\documentclass[a4paper,11pt]{article}

\usepackage[top=1in, bottom=1in, left=1in, right=1in]{geometry}
\usepackage{mcode}
\usepackage{algorithm}
\usepackage{algpseudocode}
\usepackage{array}
\usepackage{url}
\usepackage{bm}
\usepackage{graphicx}
\usepackage{color}
\usepackage[caption=false]{subfig}
\usepackage{changes}

\usepackage{hyperref}
\usepackage{datetime}

\newcommand{\sct}{\text{sct}}

\newcommand{\tot}{\text{tot}}
\newcommand{\supp}{\text{supp}}
\newcommand{\spark}{\text{spark}}
\newcommand{\rank}{\text{rank}}

\newcommand{\dB}{\text{dB}}
\newcommand{\GMMV}{\text{GMMV}}
\newcommand{\LSM}{\text{LSM}}
\graphicspath{{figures/}}

\begin{document}

\title{A Linear Method for Shape Reconstruction based on the Generalized Multiple Measurement Vectors Model}

\author{
        Shilong~Sun\thanks{S. Sun, B. J. Kooij, and A. G. Yarovoy are with the Delft University of Technology, 2628 Delft,
        The Netherlands (e-mail: shilongsun@icloud.com; B.J.Kooij@tudelft.nl; A.Yarovoy@tudelft.nl).}
        \and Bert~Jan~Kooij\footnotemark[1]
        \and Alexander~G.~Yarovoy\footnotemark[1]
        \and Tian~Jin\thanks{T. Jin is with the College of Electronic Science and Engineering,
        National University of Defense Technology, Changsha 410073, China (e-mail: tianjin@nudt.edu.cn).}}

\newdate{date}{5}{2}{2018}
\date{\displaydate{date}}

\maketitle

\maketitle

\begin{abstract}

    In this paper, a novel linear method for shape reconstruction is proposed based on the generalized multiple measurement vectors (GMMV) model. Finite difference frequency domain (FDFD) is applied to discretized Maxwell's equations, and the contrast sources are solved iteratively by exploiting the joint sparsity as a regularized constraint. Cross validation (CV) technique is used to terminate the iterations, such that the required estimation of the noise level is circumvented. The validity is demonstrated with an excitation of transverse magnetic (TM) experimental data, and it is observed that, in the aspect of focusing performance, the GMMV-based linear method outperforms the extensively used linear sampling method (LSM).  
       
\end{abstract}

\section{Introduction}\label{sec.Intro}

    Inverse electromagnetic (EM) scattering is a procedure of recovering the characteristics of unknown objects using the scattered fields probed at a number of positions. In many real applications, such as geophysical survey \cite{kuroda2007full,ernst2007full,virieux2009overview,bleistein2013mathematics}, it is of great importance to retrieve the geometrical features of a system of unknown targets. 

    For solving this problem, a wealth of reconstruction methods have been proposed over the recent decades. Due to their high efficiency, the linear focusing methods have been extensively used in real applications, among which are Kirchhoff migration \cite{schneider1978integral}, back-projection method \cite{munson1983tomographic}, time-reversal (TR) technique \cite{fink1993time,fink2000time,micolau2003dort,yavuz2005frequency,liu2007electromagnetic,yavuz2008space,fouda2012imaging,bahrami2014ultrawideband,fouda2014statistical}, and so forth. However, as is well known the imaging resolutions of the linear focusing algorithms are bound by the diffraction limit \cite{zhang2013comparison}. As a variant of TR technique, time-reversal multiple signal classification (TR MUSIC) \cite{devaney2005time,marengo2006subspace,marengo2007time,ciuonzo2015performance} is capable to achieve a resolution that can be much finer than the diffraction limit by exploiting the orthogonality of the signal and noise sub-spaces. Linear Sampling Method (LSM) \cite{colton1996simple,colton1997simple} is a non-iterative inversion technique of finding an indicator function for each position in the region of interest (ROI) by first defining a far-field (or near-field \cite{fata2004linear}) mapping operator, and then solving a linear system of equations. LSM has been proven to be effective for impenetrable scatterers, and in some cases, also applicable for dielectric scatterers \cite{arens2003linear}. As a matter of fact, LSM can also be reinterpreted, apart from very peculiar cases, as a ``synthetic focusing'' problem \cite{catapano2007simple}, and more interestingly, an extension of the MUSIC algorithm \cite{cheney2001linear}. There is another group of iterative surface-based inversion methods, which first parametrize the shape of the scatterer, then optimize the parameters by minimizing a cost functional iteratively \cite{roger1981Newton}. The drawbacks of these methods are obvious. Firstly, they require \textit{a priori} information about the position and the quantity of the scatterers. More research on this point can be found in \cite{qing2003electromagnetic,qing2004electromagnetic}. Secondly, it is intractable to deal with complicated non-convex objects. Quantitative inversion methods, such as contrast source inversion (CSI) \cite{kleinman1992modified,kleinman1993extended,kleinman1994two,van1997contrast} and (Distorted) Born iterative methods (BIM and DBIM) \cite{wang1989iterative,chew1990reconstruction,li2004three,gilmore2009comparison}, can also be used for shape reconstruction. However, it is very time consuming due to the fact that the forward scattering problem needs to be solved in every iteration.

    In this paper, a novel linear method using generalized multiple measurement vectors (GMMV) model \cite{van2010theoretical,heckel2012joint} is proposed for solving the problem of shape reconstruction. Specifically, as the objects are illuminated by EM waves from various incident angles at different frequencies, the contrast sources, i.e., the multiplication of the contrast and the total fields, are distributed in the same region with the objects. Therefore, the problem is consequently formulated as a GMMV model, and the contrast sources can be retrieved by solving multiple systems of linear equations simultaneously. In our method, the sum-of-norm of the contrast sources is used as a regularization constraint to address the ill-posedness. Finite difference frequency domain (FDFD) \cite{W.Shin2013} is used to construct the scattering operator which enables simple incorporation of complicated background media, and the spectral projected gradient method, referred to as SPGL1 \cite{BergFriedlander:2008,van2011sparse}, is selected to estimate the contrast sources by solving a sum-of-norm minimization problem. Sparse scatterer imaging has been studied in \cite{oliveri2011bayesian}, in which the single measurement vector (SMV) model was used, but the joint sparsity was not considered. The application of joint sparsity in the field of medical imaging has been reported in \cite{lee2011compressive}, which is actually a hybridization of compressive sensing (CS) \cite{candes2006robust} and MUSIC based on a so-called generalized MUSIC criterion. In the aforementioned work, sparse targets (original or equivalently transformed) and their sparsest solutions are considered, and the problem of defining the best discretization grid and target number is critical for ensuring a level of sparsity that is recoverable. Equivalence principles have been considered in \cite{bevacqua2017shape} for reconstructing the boundary of dielectric scatterers. In this paper, we use sum-of-norm as a regularization constraint and we demonstrate a regularized solution of the contrast sources is sufficient to recover the spatial profile of the non-sparse targets. In this paper, we only considered the transverse magnetic (TM) EM scattering problem, and we verified the validity of the proposed method with 2-D experimental data provided by the Institut Fresnel, France \cite{0266-5611-17-6-301,geffrin2005free} for three distinct cases -- metallic objects, dielectric objects, and a hybrid one of both. Since the noise level is unknown in real applications, cross validation (CV) technique \cite{ward2009compressed} is used to terminate the optimization process. Comparison of the inverted results indicates that the proposed method shows higher resolving ability than LSM.
    
    The remainder of the paper is organized as follows: In Section~\ref{sec.ProSta}, the problem statement is given. In Section~\ref{sec.GMMVLinMethod}, the proposed GMMV-based linear method\footnote{The GMMV-LIM package is available at \url{https://github.com/TUDsun/GMMV-LIM}.} is introduced in detail. The validation of this method with experimental data is given in Section~\ref{sec.ExpData}. Finally, Section~\ref{sec.Conclusion} ends this paper with our conclusions.

\section{Problem Statement}\label{sec.ProSta}

    For the sake of simplicity, we consider the 2-D TM-polarized EM scattering problem. A bounded, simply connected, inhomogeneous background domain $\mcD$ contains unknown objects. The domain $\mcS$ contains the sources and receivers. The sources are denoted by the subscript $p$ in which $p\in\{1,2,3...,P\}$, and the receivers are denoted by the subscript $q$ in which $q\in\{1,2,3,...,Q\}$. We use a right-handed coordinate system, and the unit vector in the invariant direction points out of the paper. 

    Assume the background is known to a reasonable accuracy beforehand, and the permeability of the background and unknown objects is a constant, denoted by $\mu_0$. The contrast corresponding to the $i$-th frequency, $\chi_i$, is defined as $\chi_i = \epsilon_i-\epsilon^{\text{bg}}_i$, where $\epsilon_i=\varepsilon-\rmi\sigma/\omega_i$ and $\epsilon^{\text{bg}}_i=\varepsilon^{\text{bg}}-\rmi\sigma^{\text{bg}}/\omega_i$ are the complex permittivity of the inversion domain with and without the presence of the targets, respectively. Here, $\varepsilon$ and $\varepsilon^{\text{bg}}$ are the permittivity of the inversion domain with and without the presence of the targets, respectively; $\sigma$ and $\sigma^{\text{bg}}$ are the conductivity of the inversion domain with and without the presence of the targets, respectively; $\omega_i$ is the $i$-th angular frequency; $\rmi$ represents the imaginary unit. The time factor used in this paper is $\exp(\rmi\omega_i t)$. For 2-D TM-polarized scattering problems, the electric field is a scalar and the scattering wave equation with respect to the scattered fields can be easily derived from Maxwell's equations, which is given by 
    \begin{equation}\label{eq.CSI.E}
        -\nabla^2 E_{p,i}^{\sct}-k_i^2 E_{p,i}^{\sct}=\omega_i^2\mu_0 J_{p,i},\quad p=1,2,\dots,P, \quad i = 1,2,\dots,I,
    \end{equation}
    where, $\nabla^2$ is the Laplace operator, $k_i=\omega_i\sqrt{\epsilon_b\mu_0}$ is the $i$-th wavenumber, $J_{p,i} = \chi_i E_{p,i}^{\tot}$ is the $p$-th contrast source at the $i$-th frequency, $E_{p,i}^{\sct}$ and $E_{p,i}^{\tot}$ are the scattered electric field and the total electric field at the $i$-th frequency, respectively. The inverse scattering problems discussed in this paper are to retrieve the geometrical features of the scatterers from a set of scattered field measurements. 

\section{The GMMV-based Linear Method}\label{sec.GMMVLinMethod}

\subsection{The GMMV Formulation}\label{subsec.GMMVFor}
    
    Following the vector form of the FDFD scheme in \cite{W.Shin2013}, we discretize the 2-D inversion space with $N$ grids and recast the scattering wave equation \eqref{eq.CSI.E} into the following matrix formalism 
    \begin{equation}\label{eq.FD-CSI.eq}
        \mA_i\ve_{p,i}^\sct=\omega_i^2\vj_{p,i}, \quad p=1,2,\dots,P, \quad i = 1,2,\dots,I,
    \end{equation}
    where $\mA_i\in\mbC^{N\times N}$ is the FDFD stiffness matrix of the $i$-th frequency, which is highly sparse; $\ve_{p,i}^\sct\in\mbC^{N}$ and $\vj_{p,i}\in\mbC^{N}$ are the scattered electric field and the contrast source in the form of a column vector, respectively. Obviously, the solution to Eq.~\eqref{eq.FD-CSI.eq} can be obtained by inverting the stiffness matrix $\mA_i$, i.e., $\ve_{p,i}^\sct=\mA_i^{-1}\omega_i^2\vj_{p,i}$. For the inverse scattering problems discussed in this paper, the scattered fields are measured with a number of receivers at specified positions, yielding the data equations given by
    \begin{equation}\label{eq.data}
        \vy_{p,i}=\bm{\Phi}_{p,i} \vj^{\text{ic}}_{p,i},\quad p=1,2,\dots,P, \quad i = 1,2,\dots,I,
    \end{equation}
    where, $\bm{\Phi}_{p,i} = \mcM^\mcS_p \mA_i^{-1}\omega_i\in\mbC^{Q\times N}$ is the sensing matrix for the measurement $\vy_{p,i}$, $\vj^{\text{ic}}_{p,i}=\omega_i\vj_{p,i}$ is the normalized contrast source proportional to the induced current $\rmi\omega_i\mu_0\vj_{p,i}$. Here, $\mcM^{\mcS}_p$ is a measurement matrix selecting the values of the $p$-th scattered field at the positions of the receivers. 

    In the rest of this subsection, a GMMV model \cite{heckel2012joint} is constructed and solved by exploiting the joint sparsity of the normalized contrast sources. In doing so, the contrast sources can be well estimated by solving a sum-of-norm minimization problem, and consequently be used to indicate the shape of the scatterers. To do so, we reformulate the data equations, Eq.~\eqref{eq.data}, as
    \begin{equation}\label{eq.linearmodel}
      \mY = \bm{\Phi} \cdot \mJ + \mU
    \end{equation}
    where 
    \begin{equation}
        \mY = 
        \begin{bmatrix}
            \vy_{1,1} & \vy_{2,1} & \dots & \vy_{P,1} & \vy_{1,2} & \dots & \vy_{P,I}
        \end{bmatrix},
    \end{equation}
    \begin{equation}
        \mJ = 
        \begin{bmatrix}
            \vj^{\text{ic}}_{1,1} & \vj^{\text{ic}}_{2,1} & \dots & \vj^{\text{ic}}_{P,1} & \vj^{\text{ic}}_{1,2} & \dots & \vj^{\text{ic}}_{P,I}
        \end{bmatrix},
    \end{equation}
    and $\bm{\Phi} \cdot \mJ$ is defined by
    \begin{equation}
        \bm{\Phi} \cdot \mJ =
        \begin{bmatrix}
            \bm{\Phi}_{1,1}\vj^{\text{ic}}_{1,1} & \bm{\Phi}_{2,1}\vj^{\text{ic}}_{2,1} & \dots & \bm{\Phi}_{P,I}\vj^{\text{ic}}_{P,I}
        \end{bmatrix},
    \end{equation}
    and, correspondingly, $\bm{\Phi}^H \cdot \mY$ is defined as
    \begin{equation}
        \bm{\Phi}^H\cdot\mY=
        \begin{bmatrix}
            \bm{\Phi}^H_{1,1}\vy^{\text{ic}}_{1,1} & \bm{\Phi}^H_{2,1}\vy^{\text{ic}}_{2,1} & \dots & \bm{\Phi}^H_{P,I}\vy^{\text{ic}}_{P,I}
        \end{bmatrix}.
    \end{equation}
    Here, $\mY\in\mbC^{Q\times PI}$ is the measurement data matrix, and the columns of $\mJ\in\mbC^{N\times PI}$ are the multiple vectors to be solved. $\mU\in\mbC^{Q\times PI}$ represents the complex additive noises satisfying certain probability distribution. It is worth noting that for single frequency inverse scattering problem, if the positions of the receivers are fixed, i.e., $\bm{\Phi}_{1,1} = \bm{\Phi}_{2,1} = \cdots=\bm{\Phi}_{Q,1}$, Eq.~\eqref{eq.linearmodel} reduces to the standard multiple measurement vectors (MMV) model \cite{van2010theoretical}.

\subsection{Guideline of the Measurement Configuration}\label{subsec.GuiMeaConf}

    Although the joint sparsity is used in this paper as a regularization constraint, an investigation on the uniqueness condition is still of much importance for two reasons: 1) It is of great interest to know how much we could benefit from a joint recovery; 2) It provides us a guideline of the measurement configuration to make the most of the joint processing. 

    According to the work of Chen and Huo \cite{chen2006theoretical} and Davies and Eldar \cite{davies2012rank}, a necessary and sufficient condition for the measurements $\mY=\mA\mX$ to uniquely determine the row sparse matrix $\mX$ is that
    \begin{equation}
        |\text{supp}(\mX)|<\frac{\spark(\mA)-1+\rank(\mX)}{2},
    \end{equation}
    where, $\supp(\mX)$ denotes the index set corresponding to non-zero rows of matrix $\mX$, $|\supp(\mX)|$ denotes the cardinality of $\supp(\mX)$, the spark of a given matrix is defined as the smallest number of the columns that are linearly dependent. Thereafter, Heckel and B{\"o}lcskei have studied the GMMV problem and showed that having different measurement matrices can lead to performance improvement over the standard MMV case \cite{heckel2012joint}. The above work about the uniqueness condition implies specifically in our method that in order to make the most of the joint processing, the column number of matrix $\mJ$ is supposed to be larger than the number of measurements, i.e., $P\times I>Q$. Moreover, with the same measurement configuration, the inversion performance can be further improved by exploiting the frequency diversity even for the case of $P>Q$. The latter is further demonstrated in Subsection~\ref{subsec.Die}.

\subsection{Spectral Projected Gradient L1 method (SPGL1)}\label{subsec.SPGL1}  

\subsubsection{GMMV basis pursuit denoise (BP\texorpdfstring{$_\sigma$}{Lg}) problem}

    Suppose the noise level is known beforehand, the approach to finding the multiple vectors is based on solving a convex optimization problem (referred to as GMMV (BP$_\sigma$) problem), which can be written as follows 
    \begin{equation}
        \text{minimize}\quad \kappa(\mJ)\quad  \text{subject to}\quad  \|\bm{\Phi} \cdot \mJ - \mY\|_F\leq \tilde\sigma,
    \end{equation}
    where, $\tilde\sigma$ represents the noise level; $\kappa(\mJ)$ is the mixed ($\alpha,\beta$)-norm defined as 
    \begin{equation}
         \|\mJ\|_{\alpha,\beta}:=\left(\sum_{n=1}^N\left\|\mJ_{n,:}^T\right\|_\beta^\alpha\right)^{1/\alpha},
    \end{equation}
    where, $\mJ_{n,:}$ denotes the $n$-th row of $\mJ$; $\|\cdot\|_\beta$ is the conventional $\beta$-norm; $(\cdot)^T$ is the transpose operator; $\|\cdot\|_F$ is the Frobenius norm which is equivalent to the mixed (2,2)-norm $\|\cdot\|_{2,2}$. In this paper, we select the mixed norm $\|\cdot\|_{1,2}$ as a regularized constraint. Although $\|\cdot\|_{1,2}$ tends to enforce the row-sparsity of the matrix $\mJ$, sparsity is not a premise for this approach. The key point is the utilization of the joint structure for improving the focusing ability. As demonstrated in the following experiments, this approach is able to image objects which are not sparse by exploitation of the frequency diversity. 

\subsubsection{Multiple GMMV Lasso (LS\texorpdfstring{$_\tau$}{Lg}) problems}

    Since it is not straightforward to solve the GMMV (BP$_\sigma$) problem, we consider the GMMV (LS$_\tau$) problem formulated as \cite{BergFriedlander:2008}
    \begin{equation}\label{eq.LSMMV}
        \text{minimize}\quad \left\|\bm{\Phi}\cdot\mJ - \mY\right\|_F\quad  \text{subject to}\quad  \left\|\mJ\right\|_{1,2} \leq \tau.
    \end{equation}
    The GMMV (LS$_\tau$) problem is equivalent to the GMMV (BP$_\sigma$) problem when $\tau = \tau_{\tilde\sigma}$. As the exact value of $\tau_{\tilde\sigma}$ is not available, a series of GMMV (LS$_\tau$) problems with different values of $\tau$ must be solved. Now let us first introduce the Pareto curve defined as follows
    \begin{equation}
        \phi_{\text{GMMV}}(\tau) = \left\|\bm{\Phi}\cdot\mJ_{\tau}^{\text{opt}} - \mY\right\|_F,
    \end{equation}
    where, $\mJ_{\tau}^{\text{opt}}$ is the optimal solution to the LS$_\tau$ problem given by Eq.~\eqref{eq.LSMMV}. When the optimal solution $\mJ_{\tau_l}^{\text{opt}}$ to the GMMV (LS$_\tau$) problem is found, $\tau_l$ is updated to $\tau_{l+1}$ by probing the Pareto curve. The searching procedure is terminated when $\phi_{\text{GMMV}}(\tau)=\tilde\sigma$. At the mean time, $\tau$ reaches $\tau_{\tilde\sigma}$. 

\subsubsection{Updating the parameter \texorpdfstring{$\tau$}{Lg}}

    As the Pareto curve is proven to be a non-increasing convex function, Newton iteration is used for updating the parameter $\tau$. 
    \begin{figure}[!t]
        \centering
        \includegraphics[height=0.45\linewidth]{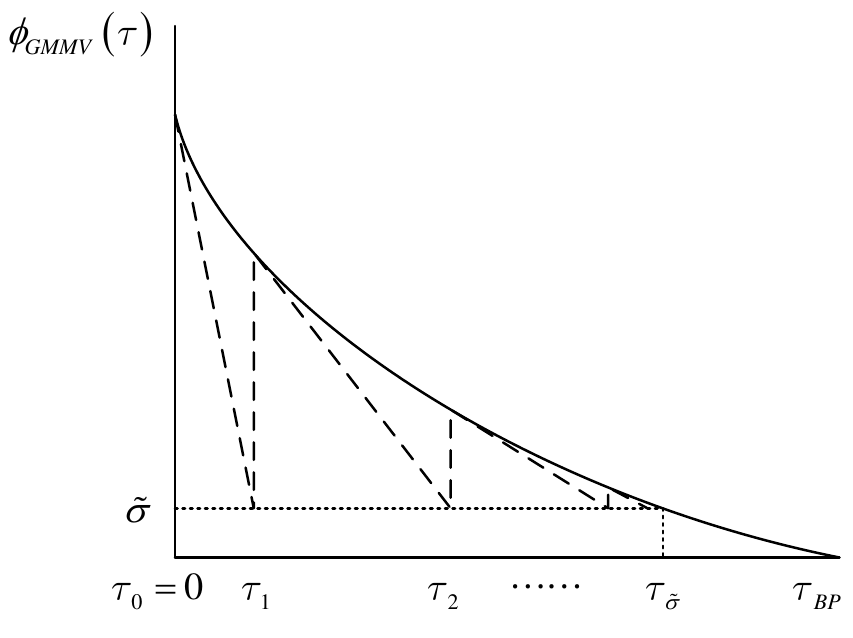}
        \caption{Probing the Pareto curve: the update of parameter $\tau$.}
        \label{fig:Pareto}
    \end{figure}
    Specifically, $\tau$ is updated by
    \begin{equation}\label{eq.tau.update}
        \tau_{l+1} = \tau_l+\frac{\tilde\sigma-\phi_{\GMMV}(\tau_l)}{\phi_{\GMMV}'(\tau_l)},
    \end{equation}
    where, 
    \begin{equation}
        \phi_{\GMMV}'(\tau_l)=-\frac{\left\|\bm{\Phi}^H\cdot(\bm{\Phi}\cdot\mJ_{\tau_l}^{\text{opt}}-\mY)\right\|_{\infty,2}}{\left\|\bm{\Phi}\cdot\mJ_{\tau_l}^{\text{opt}}-\mY\right\|_F}.
    \end{equation}
    Here, $\|\cdot\|_{\infty,2}$ is the dual norm of $\|\cdot\|_{1,2}$. The searching procedure is illustrated in Fig.~\ref{fig:Pareto}. Unless a good estimate of $\tau_{\tilde\sigma}$ can be obtained, we set $\tau_{\tilde\sigma}=0$, yielding $\phi(0) = \|\mY\|_F$ and $\phi'(0) = \|\bm{\Phi}^H\cdot\mY\|_{\infty,2}$. With Eq. \eqref{eq.tau.update}, it holds immediately that the next Newton iteration is 
    \begin{equation}
        \tau_1=\frac{\tilde\sigma-\|\mY\|_F}{\|\bm{\Phi}^H\cdot\mY\|_{\infty,2}}. 
    \end{equation} 
    We refer to \cite{BergFriedlander:2008,van2011sparse} for more details about SPGL1 and \cite{sun2017ALinearModel} for its application in inverse scattering problems.

\subsection{CV-based Modified SPGL1}\label{subsec.CVSPGL1}

    In real applications, the termination condition, $\phi_{\text{GMMV}}(\tau)=\tilde\sigma$, is not applicable, because the noise level, i.e., the parameter $\tilde\sigma$, is unknown in general. In order to deal with this problem, we modified the SPGL1 method based on the CV technique \cite{ward2009compressed}. In doing so, the estimation of the noise level can be well circumvented. 

    Specifically, we separate the original scattering matrix to a reconstruction matrix $\bm{\Phi}_{p,i,r}\in\mbC^{Q_r\times N}$ and a CV matrix $\bm{\Phi}_{p,i,\text{CV}}\in\mbC^{Q_{\text{CV}}\times N}$ with $Q = Q_r+Q_{\text{CV}}$. The measurement vector $\vy_{p,i}$ is also separated accordingly to a reconstruction measurement vector $\vy_{p,i,r}\in\mbC^{Q_r}$ and a CV measurement vector $\vy_{p,i,\text{CV}}\in\mbC^{Q_{\text{CV}}}$. The reconstruction residual and the CV residual are defined as
    \begin{equation}
        r_{\text{rec}} = \left(\sum_{i=1}^{I}\sum_{p=1}^{P}\left\|\vy_{p,i,r}-\bm{\Phi}_{p,i,r}\vj_{p,i}\right\|_2^2\right)^{1/2}
    \end{equation}
    and
    \begin{equation}
        r_{\text{CV}} = \left(\sum_{i=1}^{I}\sum_{p=1}^{P}\left\|\vy_{p,i,\text{CV}}-\bm{\Phi}_{p,i,\text{CV}}\vj_{p,i}\right\|_2^2\right)^{1/2},
    \end{equation}
    respectively. In doing so, every iteration can be viewed as two separate parts: reconstructing the contrast sources by SPGL1 and evaluating the outcome by the CV technique. The CV residual curve turns to increasing when the reconstructed signal starts to overfit the noise. The reconstructed contrast sources are selected as the output on the criterion that its CV residual is the least one. To find the least CV residual, we initialize $\tilde\sigma$ as 0 and terminate the iteration when 
    \begin{equation}\label{eq.TerCond}
        N_{\text{Iter}} > N_{\text{opt}} + \Delta N,
    \end{equation}
    is satisfied, Here, $N_{\text{Iter}}$ is the current iteration number, and $N_{\text{opt}}$ is the iteration index corresponding to the least CV residual --- the optimal solution. Namely, the CV residual is identified as the least one if the CV residual keeps increasing monotonously for $\Delta N$ iterations. In the following experimental examples, we set $\Delta N = 30$.

    Once the normalized contrast sources are obtained, one can achieve the shape of the scatterers defined as
    \begin{equation}\label{eq.GMMVimage}
        \vgamma_{\GMMV,n} = \sum_{i=1}^I\sum_{p=1}^P\left|\vj^{\text{ic}}_{p,i,n}\right|^2,\quad n=1,2,\dots,N,
    \end{equation}
    where $\vj^{\text{ic}}_{p,i,n}$ and $\vgamma_{\GMMV,n}$ represent the $n$-th element of vector $\vj^{\text{ic}}_{p,i}$ and $\vgamma_{\GMMV}$, respectively.

\section{Validation with Experimental Data}\label{sec.ExpData}

    In order to validate the proposed GMMV-based linear method, we applied it to the experimental database provided by the Remote Sensing and Microwave Experiments Team at the Institut Fresnel, France, in the years of 2001 \cite{0266-5611-17-6-301} and 2005 \cite{geffrin2005free}. Three different cases of dielectric scatterers, metallic scatterers (convex and nonconvex), and a hybrid one of both, were considered, respectively. In order to guarantee the accuracy of the FDFD scheme, the inversion domain is discretized with a grid size $\Delta^2$ satisfying
    \begin{equation}\label{eq.meshinv}
        \Delta\leq\frac{\min\{\lambda_i\}}{15},\quad i = 1,2,\dots,I, 
    \end{equation} 
    where, $\lambda_i$ is the wavelength of the $i$-th frequency.

    We have also processed the same data by LSM for comparison. Since the background of the experiments is free space and only TM wave is considered, the LSM method consists in solving the integral equation of the indicator function $g_i(\vx_s,\vx_t)$ at the $i$-th frequency
    \begin{equation}\label{eq.LSM}
        \int E_i(\vx_r,\vx_t)g_i(\vx_s,\vx_t)d\vx_t = \frac{\omega_i\mu_0}{4}H_0^{(1)}(-k_i\|\vx_s-\vx_r\|_2),
    \end{equation}
    where, $E_i(\vx_r,\vx_t)$ is the scattered field probed at $\vx_r$ corresponding to the transmitter at $\vx_t$ and the $i$-th frequency. Here, $\vx_s$ is the sampling point in the inversion domain, $H_0^{(1)}(\cdot)$ is the Hankel function of the first kind, $k_i$ is the wavenumber of the $i$-th frequency. Eq.~\eqref{eq.LSM} can be reformulated as a set of systems of linear equations
    \begin{equation}\label{eq.LSMEq}
        \mF_i\vg_{i,\vx_s}=\vf_{i,\vx_s}, \quad i = 1,2,\dots,I,
    \end{equation}
    where, $\mF_i$ is the measurement data matrix, $\vg_{i,\vx_s}$ is the indicator function of the sampling point $\vx_s$ in the form of a column vector, $\vf_{i,\vx_s}$ is the right side of Eq.~\eqref{eq.LSM} in the form of a column vector, the index $i$ represents the $i$-th frequency. Following the same approach of solving Eq.~\eqref{eq.LSMEq} in \cite{catapano2007simple,crocco2012linear}, the indicator function $\vg_{i,\vx_s}$ is sought to be
    \begin{equation}
        \|\vg_{i,\vx_s}\|^2 = \sum_{d=1}^D\left(\frac{s_{i,d}}{s_{i,d}^2+a_i^2}\right)^2\left|\vu_d^H\vf_{i,\vx_s}\right|^2,
    \end{equation}
    where, $s_{i,d}$ represents the singular value of matrix $\mF_i$ corresponding to the singular vector $\vu_d$; $(\cdot)^H$ is the conjugate transpose operator; $D=\min\{P,Q\}$; $a_i=0.01\times\underset{d}{\max}\{s_{i,d}\}$. The shape of the scatterers is defined by 
    \begin{equation}\label{eq.LSMimage}
        \vgamma_{\LSM}(\vx_s) = \frac{1}{\|\vg^{\text{MF}}_{\vx_s}\|^2},
    \end{equation}
    where, $\|\vg^{\text{MF}}_{\vx_s}\|^2$ is a multi-frequency modified indicator defined as the average of the normalized modified ones computed at each frequency \cite{catapano2008improved}
    \begin{equation}
        \left\|\vg^{\text{MF}}_{\vx_s}\right\|^2 = \frac{1}{I}\sum_{i=1}^I\frac{\left\|\vg_{i,\vx_s}\right\|^2}{\underset{\vx_s\in\mcD}{\max}\left(\|\vg_{i,\vx_s}\|^2\right)}.
    \end{equation}

    It is worth mentioning that both the normalized contrast sources and the indicator functions are proportional to the amplitude of the electric field. According to the definitions in Eq.~\eqref{eq.GMMVimage} and Eq.~\eqref{eq.LSMimage}, $\vgamma_{\GMMV}$ and $\vgamma_{\LSM}$ are proportional and inversely proportional to the power of electric fields, respectively. Therefore, the dB scaling shown in the following examples is defined as follows
    \begin{equation}
        \vgamma_{\dB}=10\times \log_{10}\left(\frac{\vgamma}{\max\{\vgamma\}}\right).
    \end{equation}

\subsection{Dielectric Scatterers}\label{subsec.Die}

\subsubsection{Example 1}

    \begin{figure}[!t]
        \centering
        \includegraphics[height=0.45\linewidth]{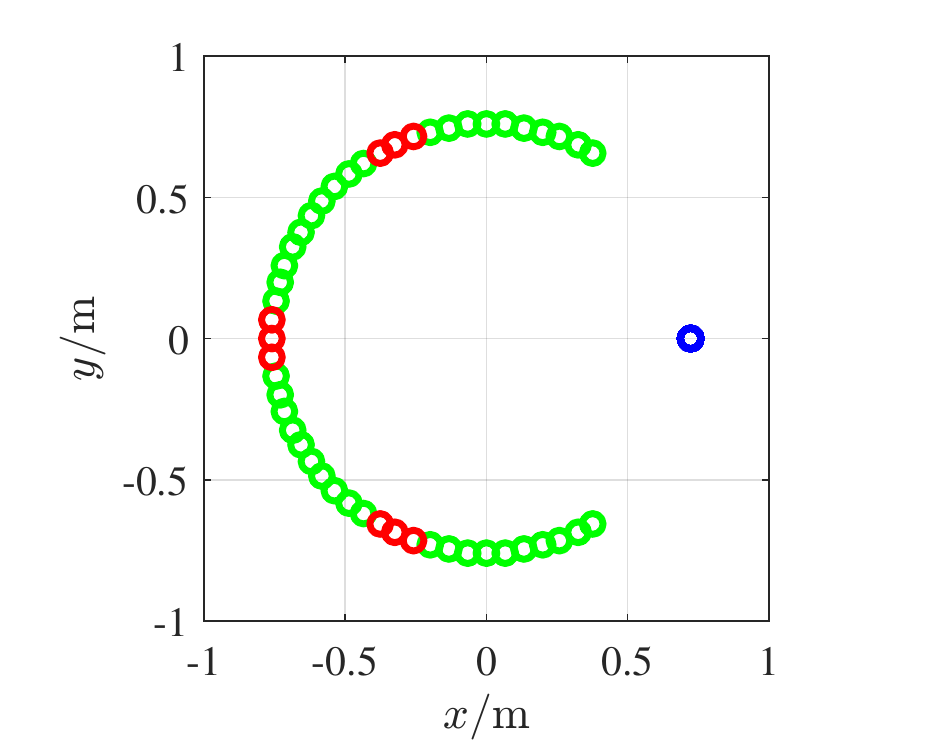}
        \caption{Measurement configuration of the data-sets: \textit{twodielTM\_8f}, \textit{rectTM\_dece}, and \textit{uTM\_shaped}. Blue: emitter; Green: reconstruction measurements; Red: CV measurements.}
        \label{fig:DieConf}
    \end{figure}

    \begin{figure}[!t]
        \centering
        \subfloat[]{\includegraphics[height=0.375\linewidth] {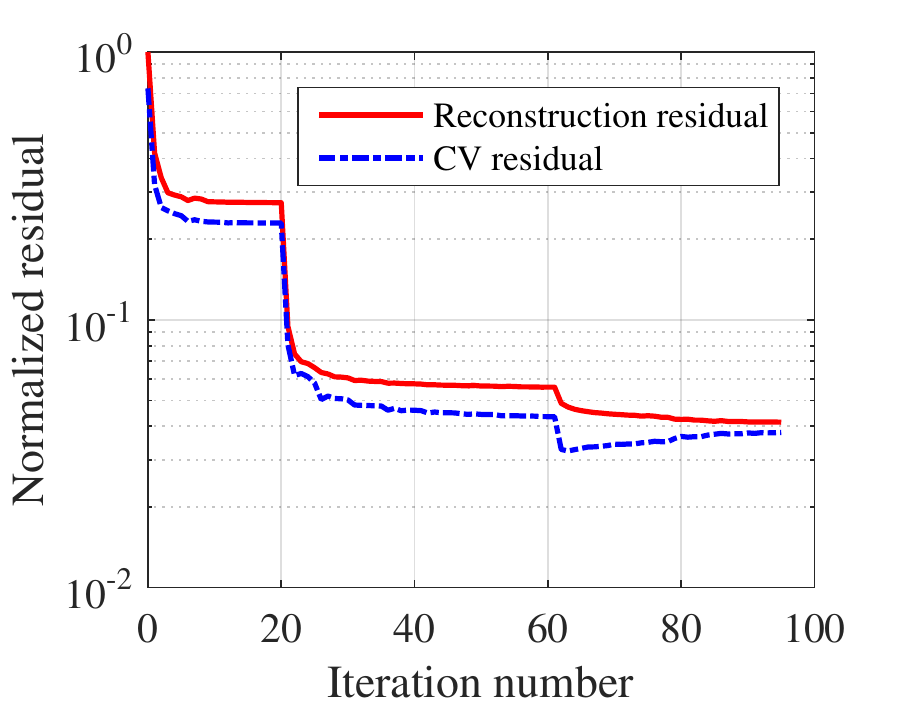}}%
        \subfloat[]{\includegraphics[height=0.375\linewidth] {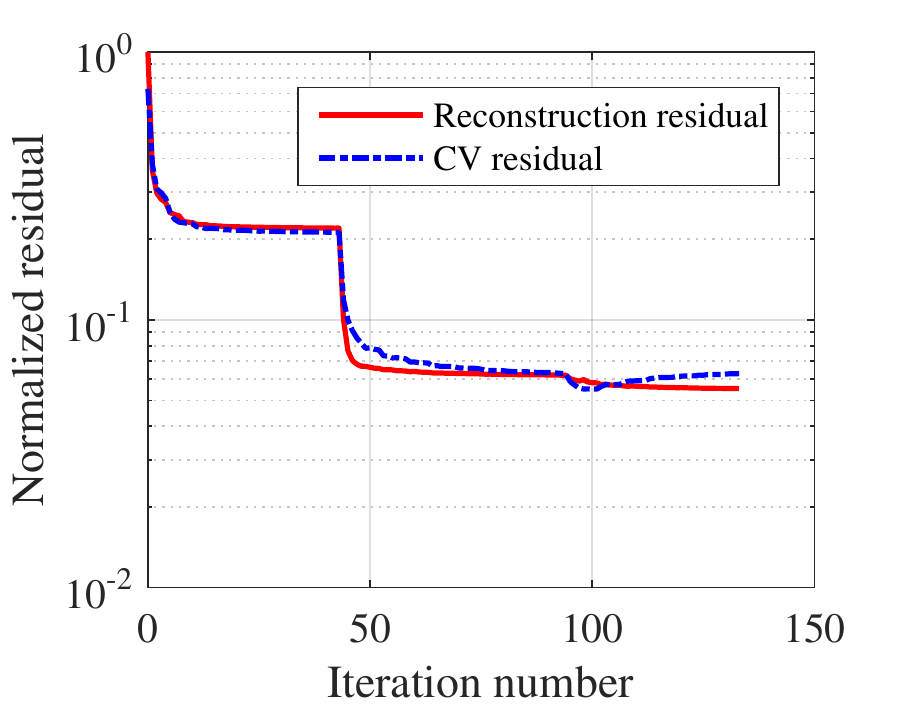}}
        \caption{Normalized reconstruction residual curve and CV residual curve in Example 1, Subsection~\ref{subsec.Die}. (a) Reconstruction with single frequency at 4 GHz; (b) Reconstruction with multiple frequencies at 2 GHz, 4 GHz, 6 GHz, and 8 GHz.}
        \label{fig:Die1CV}
    \end{figure}
    \begin{figure}[!t]
        \centering
        \subfloat[]{\includegraphics[height=0.35\linewidth] {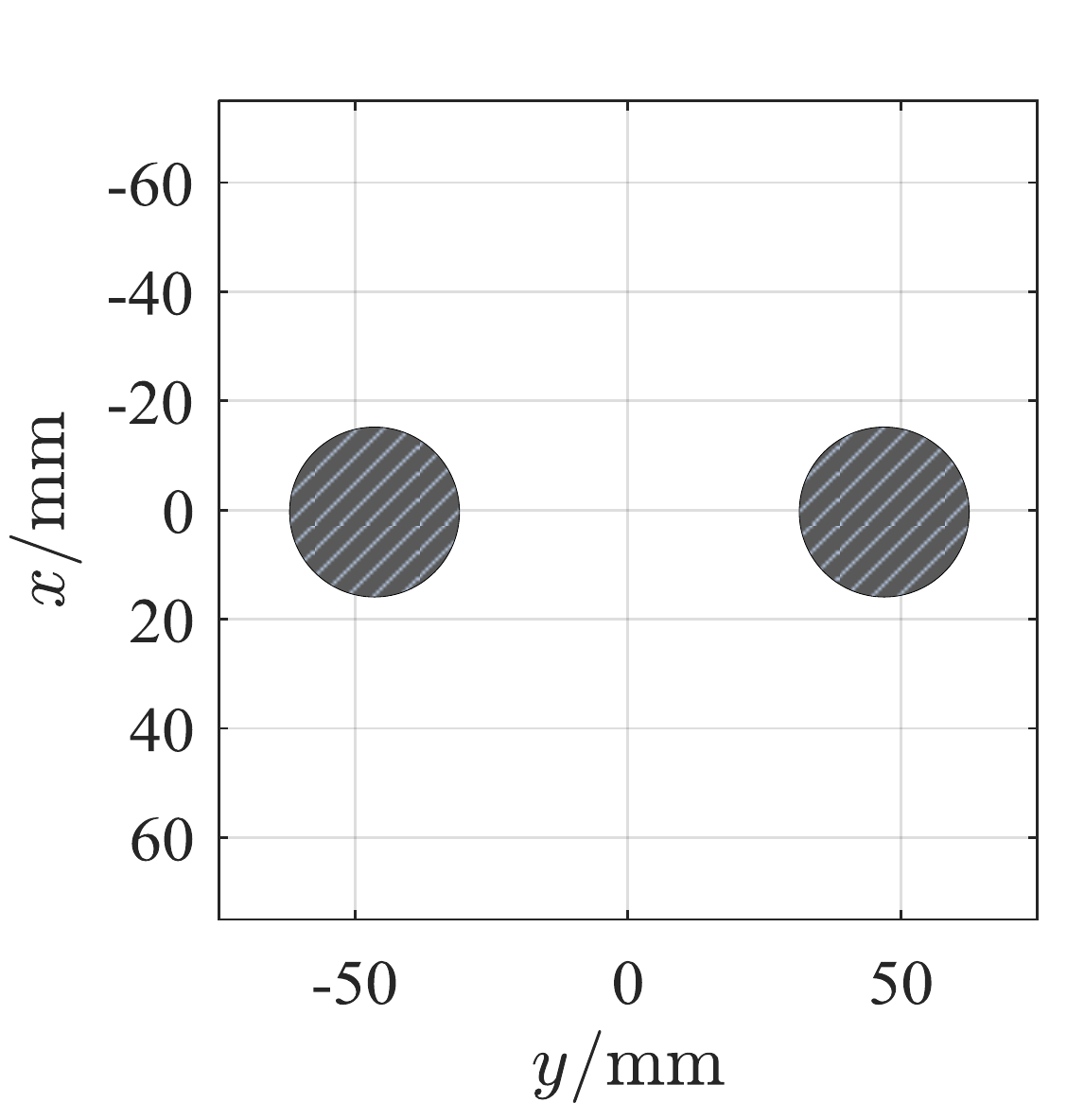}}\\
        \makebox[\columnwidth]{
        \subfloat[]{\includegraphics[height=0.35\linewidth] {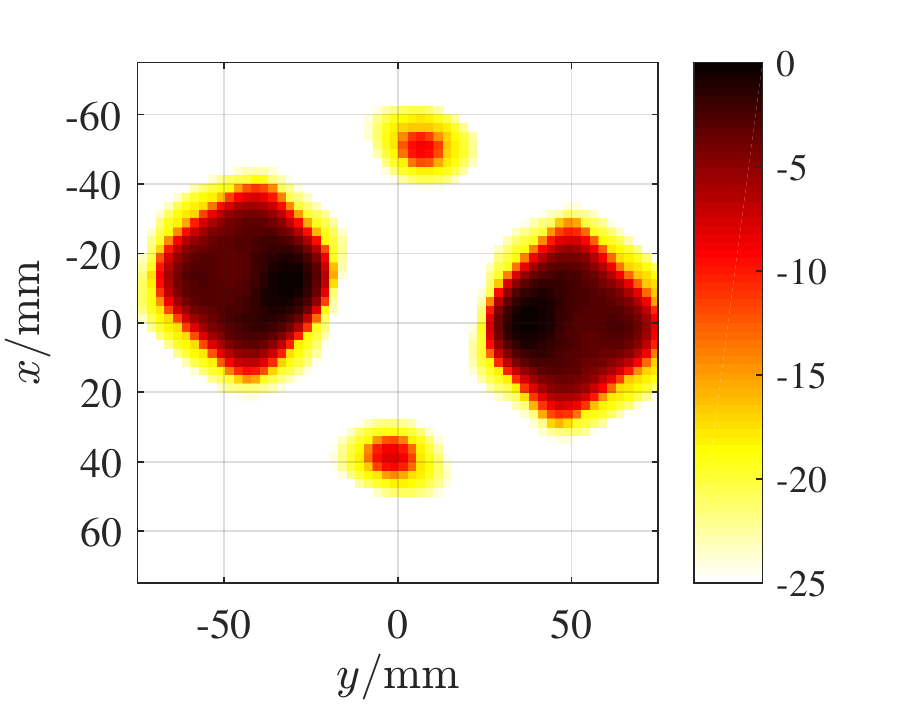}}%
        \subfloat[]{\includegraphics[height=0.35\linewidth] {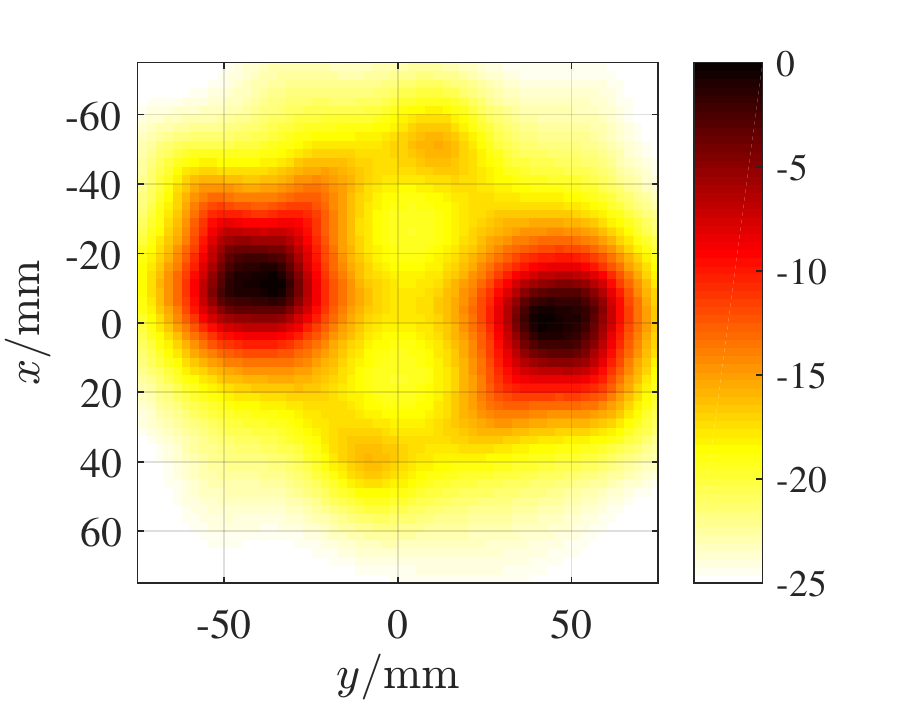}}}\\
        \makebox[\columnwidth]{
        \subfloat[]{\includegraphics[height=0.35\linewidth] {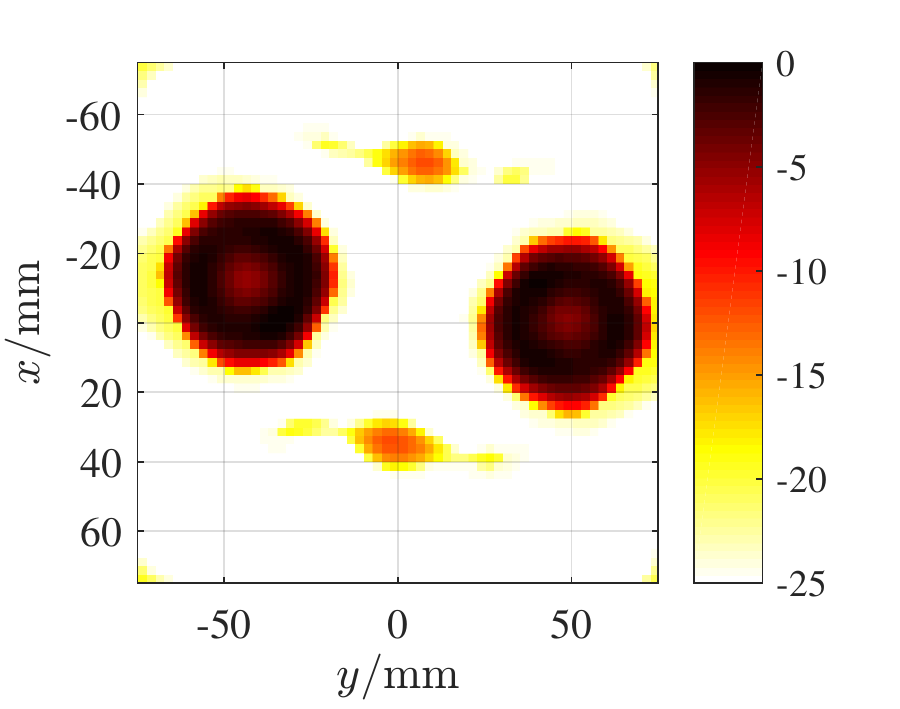}}%
        \subfloat[]{\includegraphics[height=0.35\linewidth] {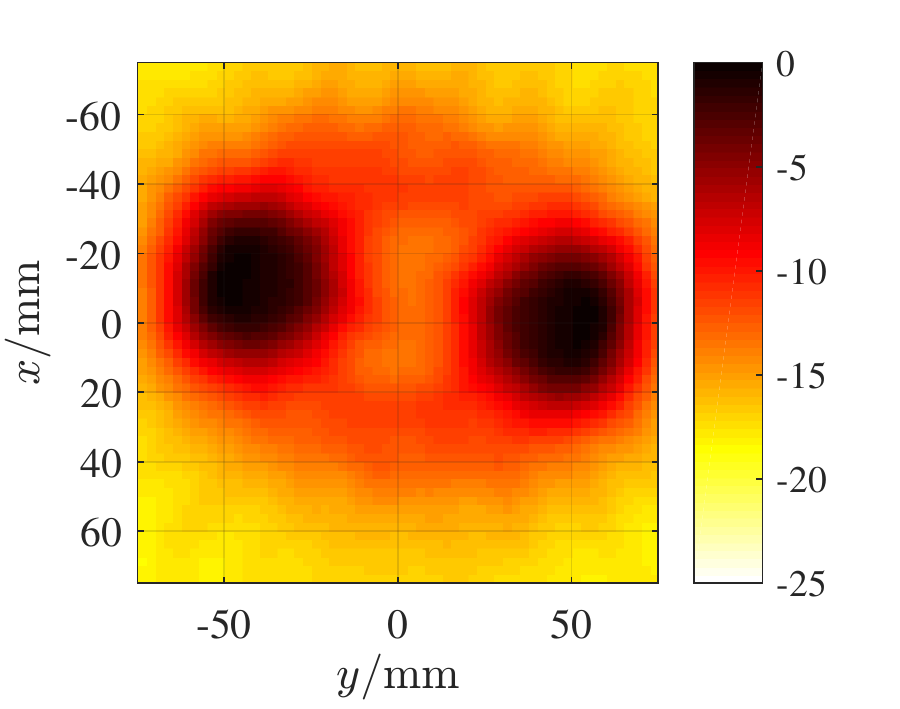}}}
        \caption{Scatterer geometry and its reconstructed shapes for Example 1 in Subsection~\ref{subsec.Die}: Scatterer geometry (a); The scatterer shape reconstructed by GMMV (b) and LSM (c) by processing the 4GHz data; The scatterer shape reconstructed by GMMV (d) and LSM (e) by processing the multiple frequency data at 2 GHz, 4 GHz, 6 GHz, and 8 GHz.}
        \label{fig:Die1}
    \end{figure}

    In the first example, we consider the \textit{twodielTM\_8f} data-set provided in the first opus of the Institut Fresnel’s database \cite{0266-5611-17-6-301}. The targets consist of two identical circular cylinders, which are shown in Fig.~\ref{fig:Die1}~(a). All the cylinders have radius of 1.5 cm and relative permittivity $3\pm 0.3$. The emitter is placed at a fixed position on the circular rail, while a receiver is rotating around the center point of the vertical cylindrical target. The targets rotated from 0$^{\circ}$ to 350$^{\circ}$ in steps of 10$^{\circ}$ with a radius of $720\pm3$ mm, and the receiver rotated from 60$^{\circ}$ to 300$^{\circ}$ in steps of 5$^{\circ}$ with a radius of $760\pm3$ mm. Namely, we have $49$ $\times$ $36$ measurement data at each frequency when all the measurements are finished. The measurement configuration is shown in Fig.~\ref{fig:DieConf}, from which we can see 9 red circles which represents the CV measurements and 40 green ones which represents the reconstruction measurements. The inversion domain is restricted to [$-75$, 75] $\times$ [$-75$, 75] mm$^2$, and the size of the discretization grids is 2.5 $\times$ 2.5 mm$^2$.  

    Let us first process the single frequency data at 4 GHz by the GMMV-based linear method and the LSM method. The data matrix $\mF_i$ for the LSM is a 72 $\times$ 36 matrix in which the data entries that are not available are replaced with zeros. The reconstruction residual curve and the CV residual curve are shown in Fig.~\ref{fig:Die1CV}~(a), from which we see the CV residual decreases before the 52-nd iteration and starts to increase thereafter. The solutions at the turning point correspond to the optimal ones. In addition, the reconstruction residual corresponding to the turning point gives an estimation of the noise level $\tilde\sigma \approx 0.05\|\mY\|_F$. Fig.~\ref{fig:Die1}~(b) and Fig.~\ref{fig:Die1}~(c) show the images achieved by the two methods at 4 GHz in a dynamic range of [$-25$, $0$] dB. As we can see the GMMV image is more clear than the LSM image. However, there is obvious shape distortion in the former. Note that $Q = 49$, $P=36$ and $I = 1$, we have $P\times I<Q$. Recalling the guideline of the measurement configuration discussed in Subsection~\ref{subsec.GuiMeaConf}, the reconstruction performance can be further improved via exploiting the frequency diversity. It is worth mentioning that an obvious position mismatch of the true objects and the reconstructed result can be observed. The reason is very likely to be the minor displacement and tilt occurred in the placement of the objects while doing this measurement, because the same phenomenon can be observed as well in the inverted results reported in \cite{bloemenkamp2001inversion}.   

    Now let us process the data at 2 GHz, 4 GHz, 6 GHz, and 8 GHz, simultaneously. The residual curves are shown in Fig.~\ref{fig:Die1CV}~(b) and the reconstructed images are shown in Fig.~\ref{fig:Die1}~(d) and Fig.~\ref{fig:Die1}~(e). By comparison of Fig.~\ref{fig:Die1}~(b) and Fig.~\ref{fig:Die1}~(d), one can see that the reconstruction performance of the proposed GMMV-based linear method is improved by exploiting the frequency diversity. One can also observe that the GMMV-based linear method achieves lower sidelobes than LSM in the case of dielectric scatterers.

\subsubsection{Example 2}

    \begin{figure}[!t]
        \centering
        \includegraphics[height=0.45\linewidth]{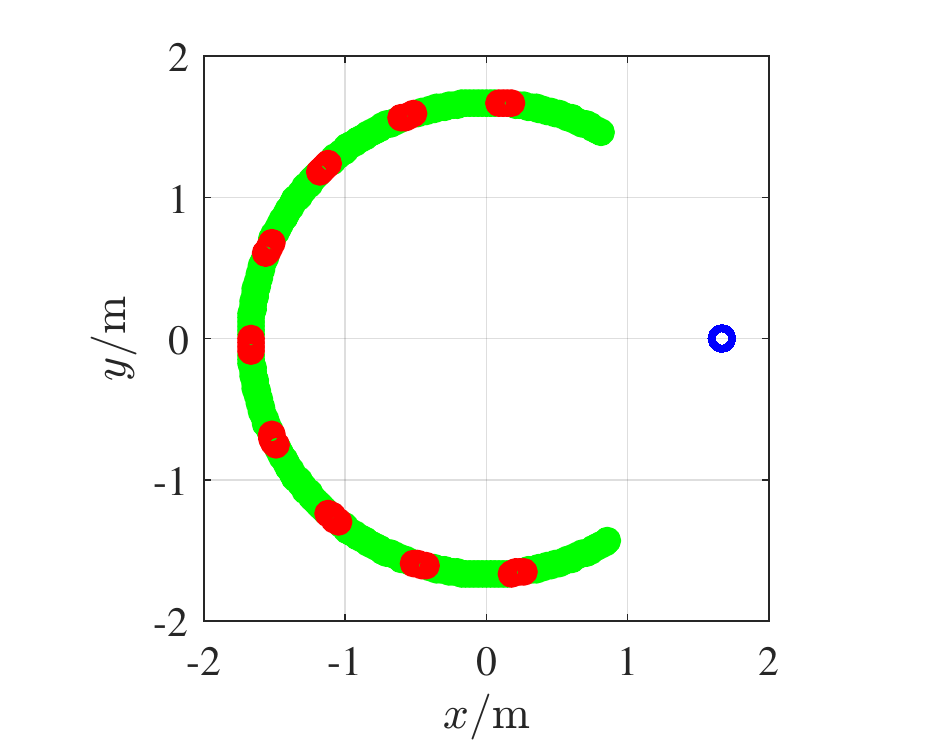}
        \caption{Measurement configuration of the data-sets: \textit{FoamDieIntTM} and \textit{FoamMetExtTM}. Blue: emitter; Green: reconstruction measurements; Red: CV measurements.}
        \label{fig:DieMetConf}
    \end{figure}

    \begin{figure}[!t]
        \centering
        \includegraphics[height=0.375\linewidth]{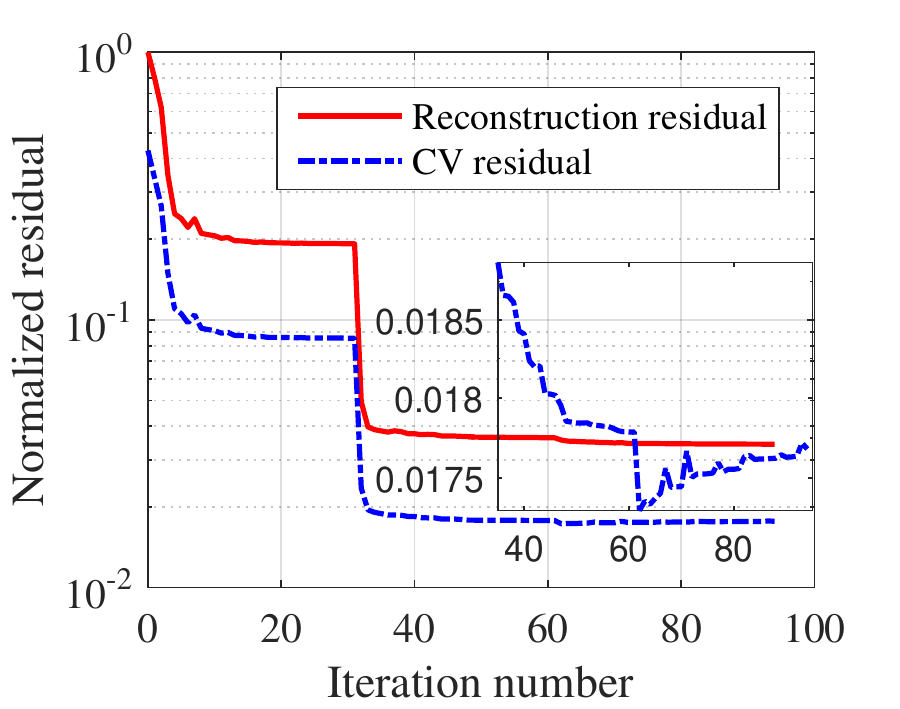}
        \caption{Normalized reconstruction residual curve and CV residual curve in Example 2, Subsection~\ref{subsec.Die}. The \textit{FoamDieIntTM} data at 2 GHz, 4 GHz, 6 GHz, 8 GHz, and 10 GHz are jointly processed.}
        \label{fig:Die2Err}
    \end{figure}
    
    \begin{figure}[!t]
        \centering
        \subfloat[]{\includegraphics[height=0.35\linewidth] {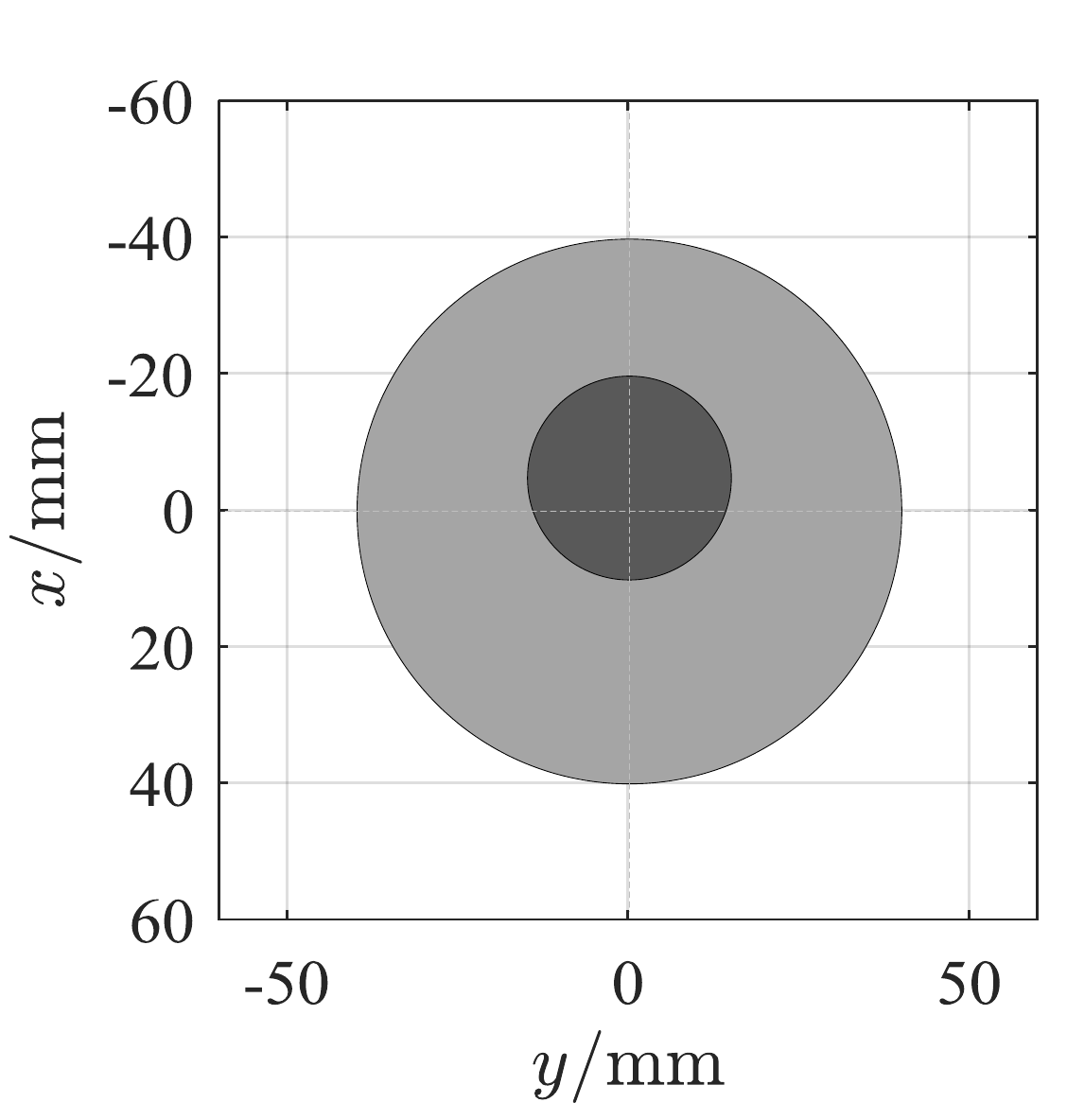}}\\
        \makebox[\columnwidth]{
        \subfloat[]{\includegraphics[height=0.35\linewidth] {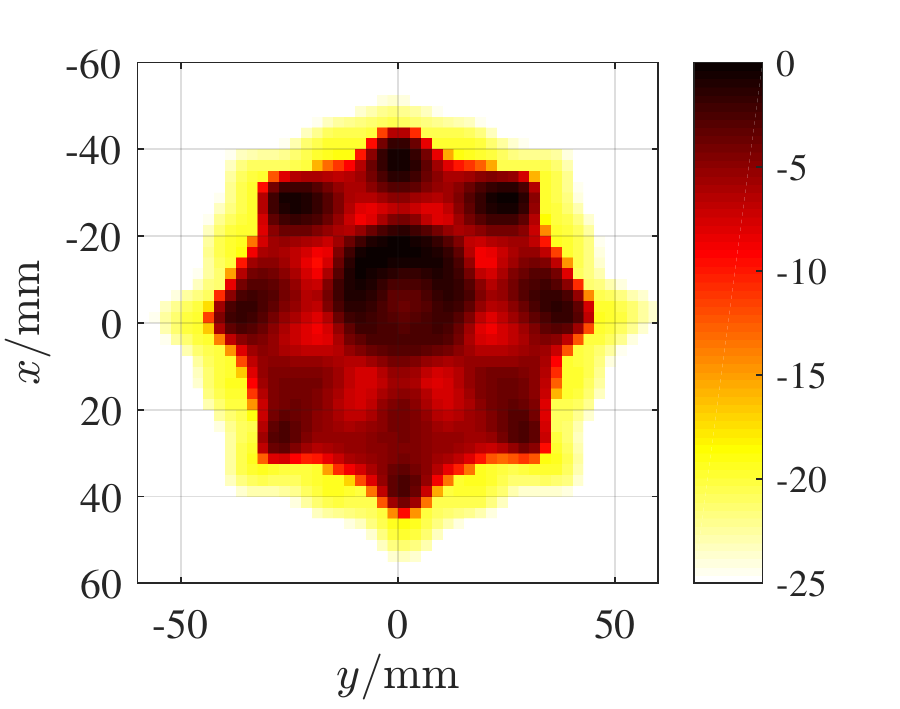}}%
        \subfloat[]{\includegraphics[height=0.35\linewidth] {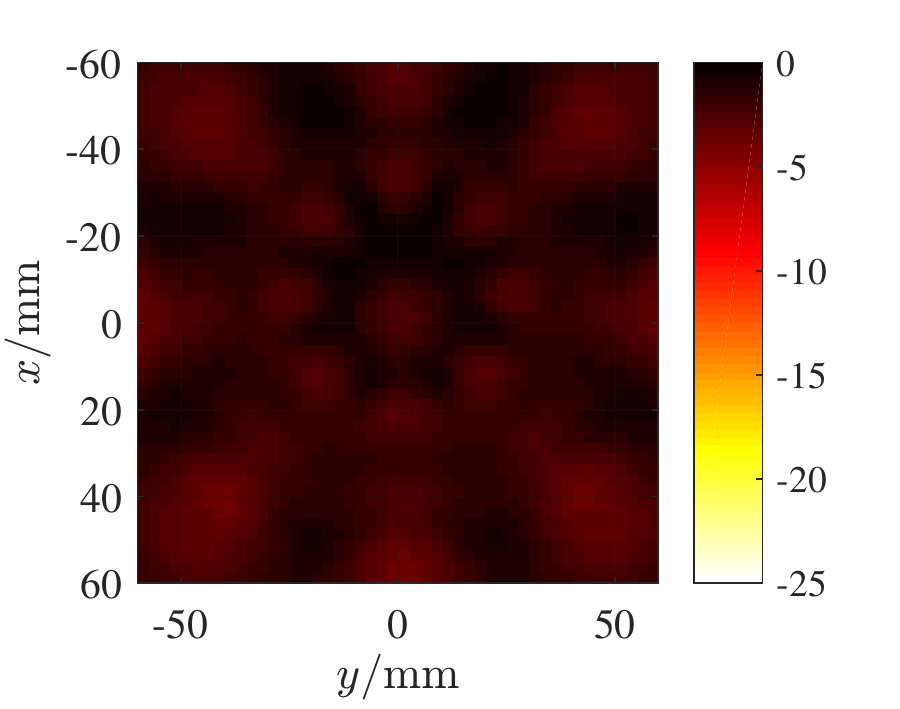}}}
        \caption{Scatterer geometry and its reconstructed shapes for the \textit{FoamDieIntTM} data-set at 2 GHz, 4 GHz, 6 GHz, 8 GHz, and 10 GHz: Scatterer geometry (a); The scatterer shape reconstructed by GMMV (b) and LSM (c).}
        \label{fig:Die2}.   
    \end{figure}

    In the second example, we consider the \textit{FoamDieIntTM} data-set provided in the second opus of the Institut Fresnel’s database. The targets consist of a circular dielectric cylinder with a diameter of 30 mm embedded in another circular dielectric cylinder with a diameter of 80 mm. The smaller cylinder has a relative permittivity value of $\varepsilon_r = 3\pm 0.3$, while the larger cylinder has a relative permittivity value of $\varepsilon_r = 1.45\pm 0.15$. Fig.~\ref{fig:Die2}~(a) shows the true objects, and we refer to \cite{geffrin2005free} for more description of the targets. The experiment is carried out in 2005, in which the receiver stays in the azimuthal plane ($xoy$) and is rotated along two-thirds of a circle from 60$^\circ$ to 300$^\circ$ with the angular step being 1$^\circ$. The source antenna stays at the fixed location ($\theta = 0^\circ$), and the object is rotated to obtain different illumination incidences from 0$^\circ$ to 315$^\circ$ with angular step of 45$^\circ$. Namely, we have $241 \times 8$ measurements at each frequency. The distance from the transmitter/receiver to the centre of the targets has increased to 1.67 m. The measurement configuration is shown in Fig.~\ref{fig:DieMetConf}, in which the blue one represents the emitter, the $4 \times 9$ red ones represent the CV measurements, and the green ones are the reconstruction measurements.

    The inversion domain is restricted to [$-60$, $60$] $\times$ [$-60$, $60$] mm$^2$, and the discretization grid size is 2.5 $\times$ 2.5 mm$^2$. Let us process the multi-frequency data at 2 GHz, 4 GHz, 6 GHz, 8 GHz, and 10 GHz simultaneously by the GMMV-based linear method and the LSM method, respectively. The data matrix $\mF_i$ for LSM is a $360 \times 8$ matrix in which the data entries that are not available are replaced with zeros. The reconstruction residual curve and the CV residual curve are shown in Fig.~\ref{fig:Die2Err}, from which we see the CV residual decreases during the first 62 iterations and starts to increase thereafter. Fig.~\ref{fig:Die2}~(b) and Fig.~\ref{fig:Die2}~(c) show the reconstructed images by the GMMV-based linear method and LSM, respectively. One can observe that the profile of the objects is reconstructed by the proposed method with high resolution, while in the LSM image the objects cannot be distinguished at all.  

\subsection{Metallic Scatterers}\label{subsec.Met}

    \begin{figure}[!t]
        \centering
        \subfloat[]{\includegraphics[height=0.375\linewidth] {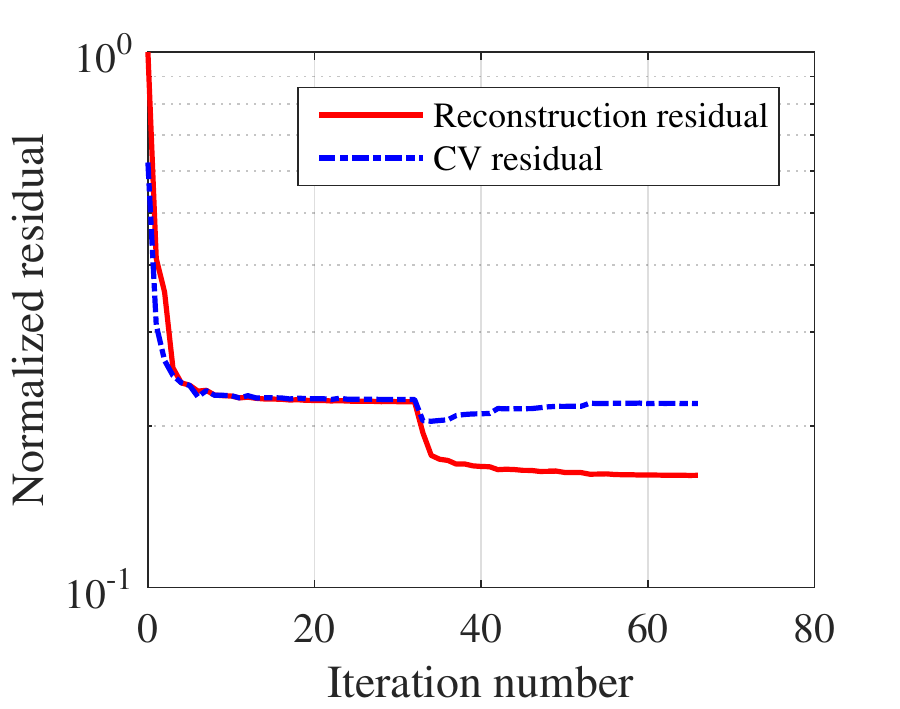}}%
        \subfloat[]{\includegraphics[height=0.375\linewidth] {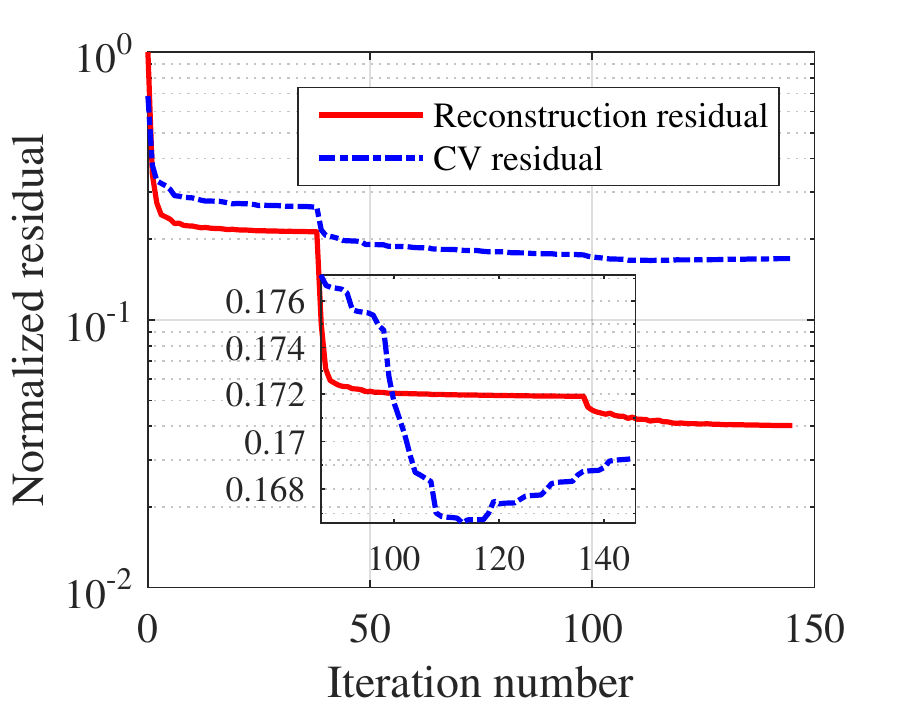}}
        \caption{Normalized reconstruction residual curve and the CV residual curve in Subsection~\ref{subsec.Met}. (a) The rectangular metallic cylinder at 10 GHz, 12 GHz, 14 GHz, and 16 GHz; (b) The ``U-shaped'' metallic cylinder at 4 GHz, 8 GHz, 12 GHz, and 16 GHz.}
        \label{fig:MetCV}
    \end{figure}  
    \begin{figure}[!t]
        \centering
        \subfloat[]{\includegraphics[height=0.35\linewidth] {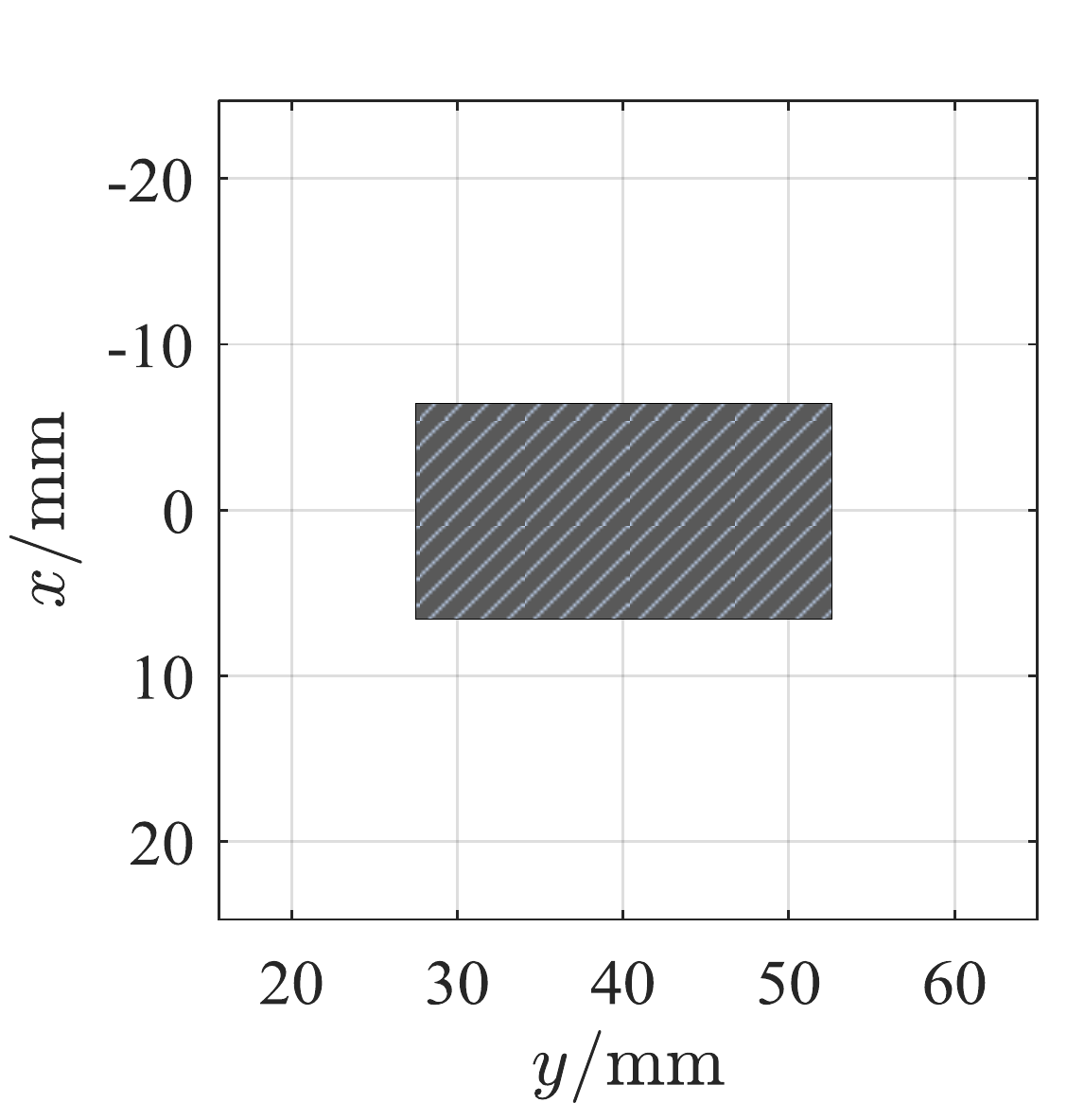}}\\
        \makebox[\columnwidth]{
        \subfloat[]{\includegraphics[height=0.35\linewidth] {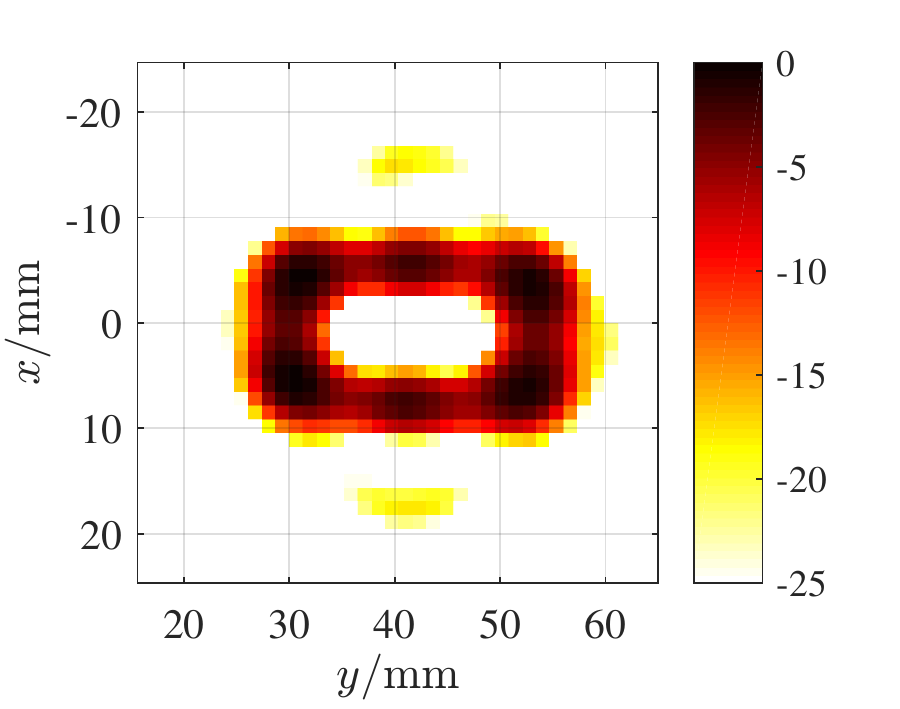}}%
        \subfloat[]{\includegraphics[height=0.35\linewidth] {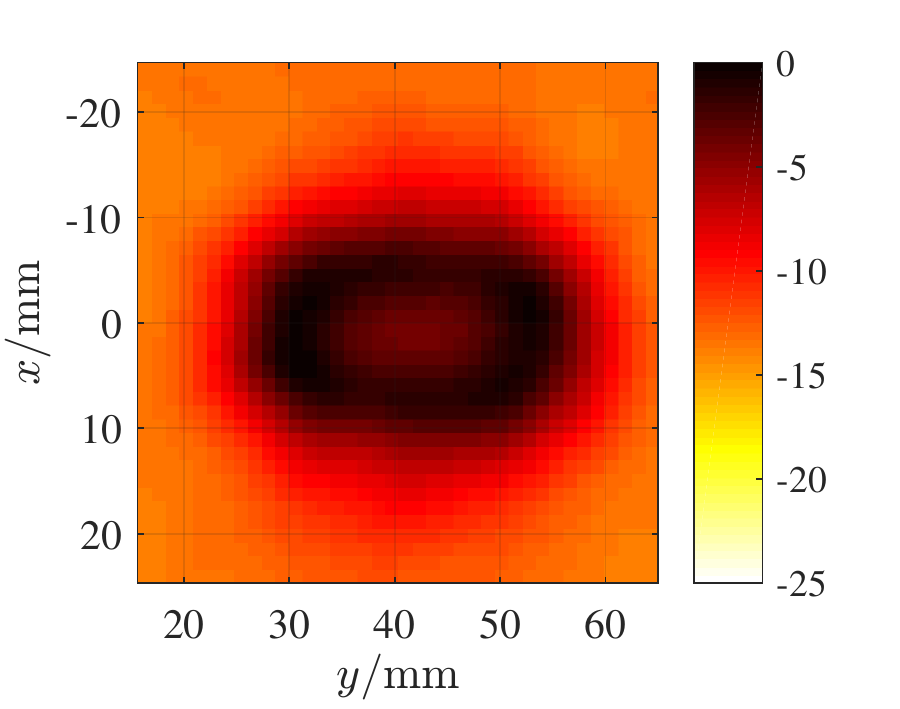}}}
        \caption{Scatterer geometry and its reconstructed shapes for the rectangular metallic cylinder obtained by processing the multiple frequency data at 10 GHz, 12 GHz, 14 GHz, and 16 GHz. : Scatterer geometry (a); The scatterer shape reconstructed by GMMV (b) and LSM (c).}
        \label{fig:Met1}.   
    \end{figure}  
    \begin{figure}[!t]
        \centering
        \subfloat[]{\includegraphics[height=0.35\linewidth] {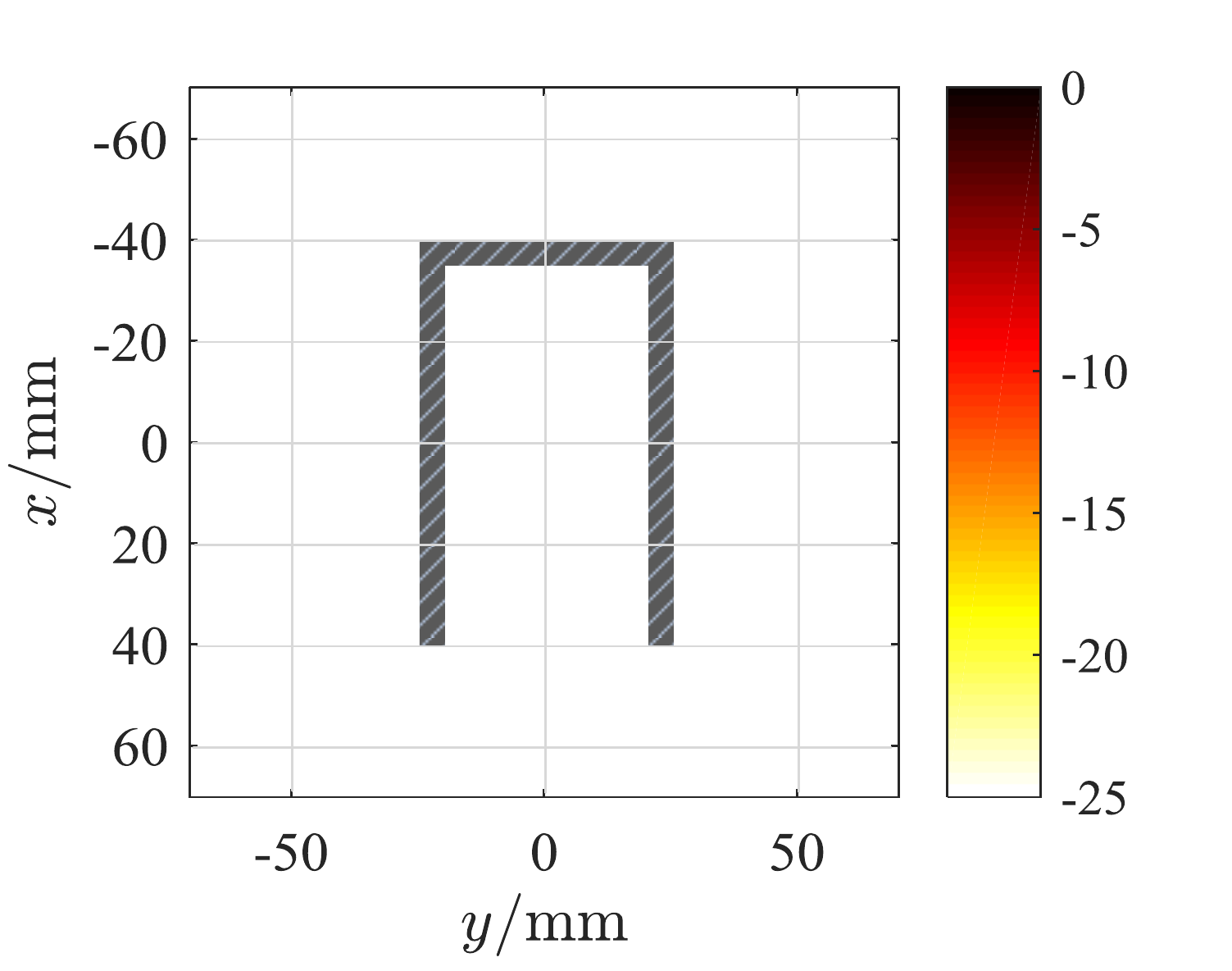}}\\
        \makebox[\columnwidth]{
        \subfloat[]{\includegraphics[height=0.35\linewidth] {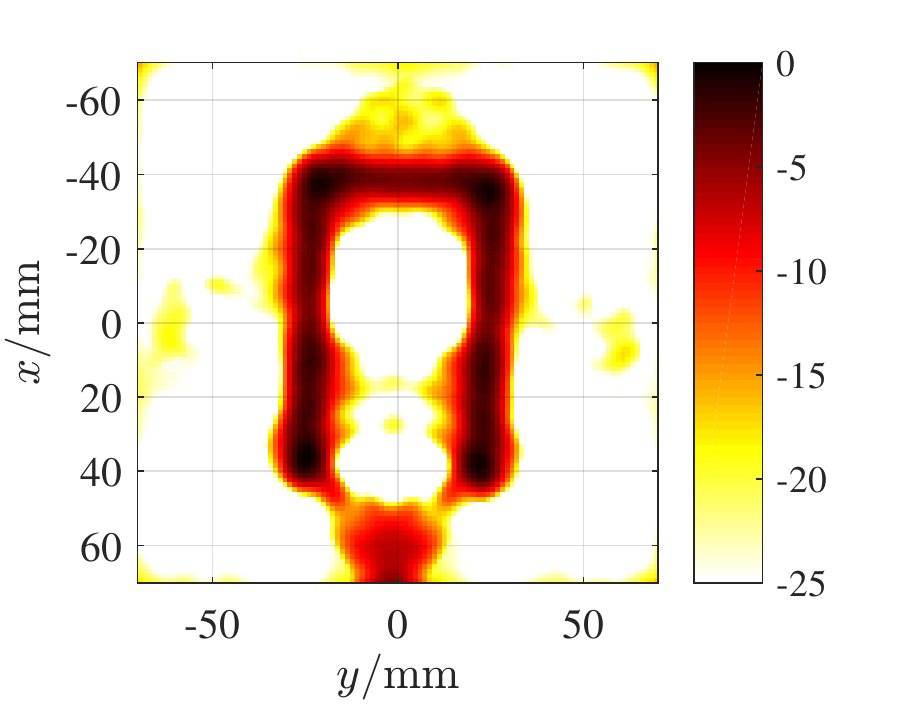}}%
        \subfloat[]{\includegraphics[height=0.35\linewidth] {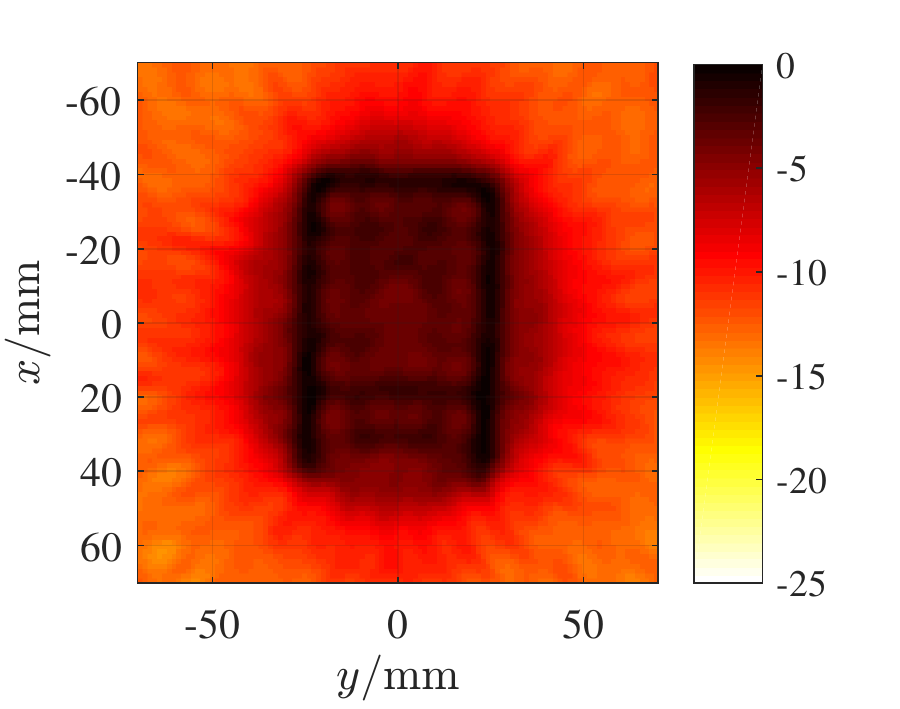}}}
        \caption{Scatterer geometry and its reconstructed shapes for the ``U-shaped'' metallic cylinder obtained by processing the multiple frequency data at 4 GHz, 8 GHz, 12 GHz, and 16 GHz: Scatterer geometry (a); The scatterer shape reconstructed by GMMV (b) and LSM (c).}
        \label{fig:Met2}.   
    \end{figure}  

    In this subsection, we applied the proposed method to the \textit{rectTM\_dece} and \textit{uTM\_shaped} data-sets provided in the first opus of the Institut Fresnel’s database \cite{0266-5611-17-6-301}, which correspond to a convex scatterer -- a rectangular metallic cylinder, and a non-convex scatterer -- a ``U-shaped'' metallic cylinder, respectively. The dimensions of the rectangular cross section are 24.5 $\times$ 12.7 mm$^2$, while those of the ``U-shaped'' cylinder are about 80 $\times$ 50 mm$^2$. The measurement configuration is same with that in Subsection~\ref{subsec.Die}. More details about the description of the targets can be found in \cite{0266-5611-17-6-301}. 

    For the rectangular cylinder, the inversion domain is restricted to [$-25$, 25] $\times$ [15, 65] mm$^2$ and the multiple frequency data at 10 GHz, 12 GHz, 14 GHz, and 16 GHz are processed simultaneously. While for the larger ``U-shaped'' cylinder, the inversion domain is restricted to [$-70$, 70] $\times$ [$-70$, 70] mm$^2$ and the multiple frequency data at 4 GHz, 8 GHz, 12 GHz, and 16 GHz are processed simultaneously. The size of the discretization grids is 1.3 $\times$ 1.3 mm$^2$. Fig.~\ref{fig:MetCV}~(a,b) give the residual curves and the reconstructed images are shown in Fig.~\ref{fig:Met1} and Fig.~\ref{fig:Met2}, respectively, from which we can see that the focusing performance of LSM is poor in the rectangular cylinder case, and it is even worse in retrieving the shape of the non-convex ``U-shaped'' cylinder, while the rectangular shape and ``U'' shape are well reconstructed by the proposed GMMV-based linear method, indicating that the latter shows higher resolving ability than the former in both the convex metallic target case and the non-convex metallic target case. 

\subsection{Hybrid Scatterers}\label{subsec.DieMet}

    \begin{figure}[!t]
        \centering
        \includegraphics[height=0.375\linewidth]{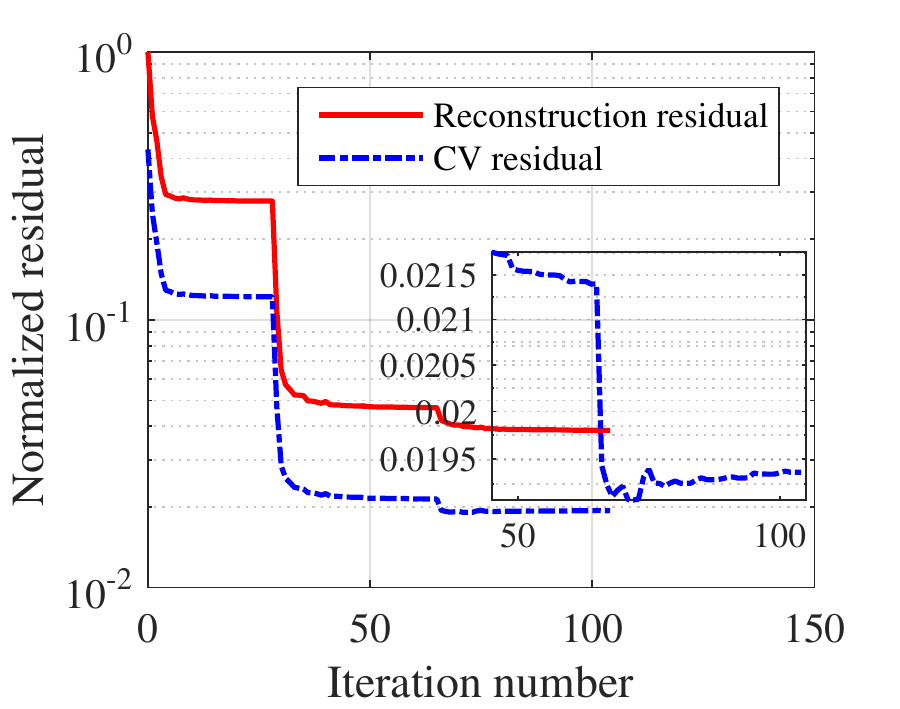}
        \caption{Normalized reconstruction residual curve and the CV residual curve in Subsection~\ref{subsec.DieMet}. The \textit{FoamMetExtTM} data-set at 2 GHz, 3 GHz, \dots, 8 GHz is processed.}
        \label{fig:DieMetCV}
    \end{figure}
    
    \begin{figure}[!t]
        \centering
        \subfloat[]{\includegraphics[height=0.35\linewidth] {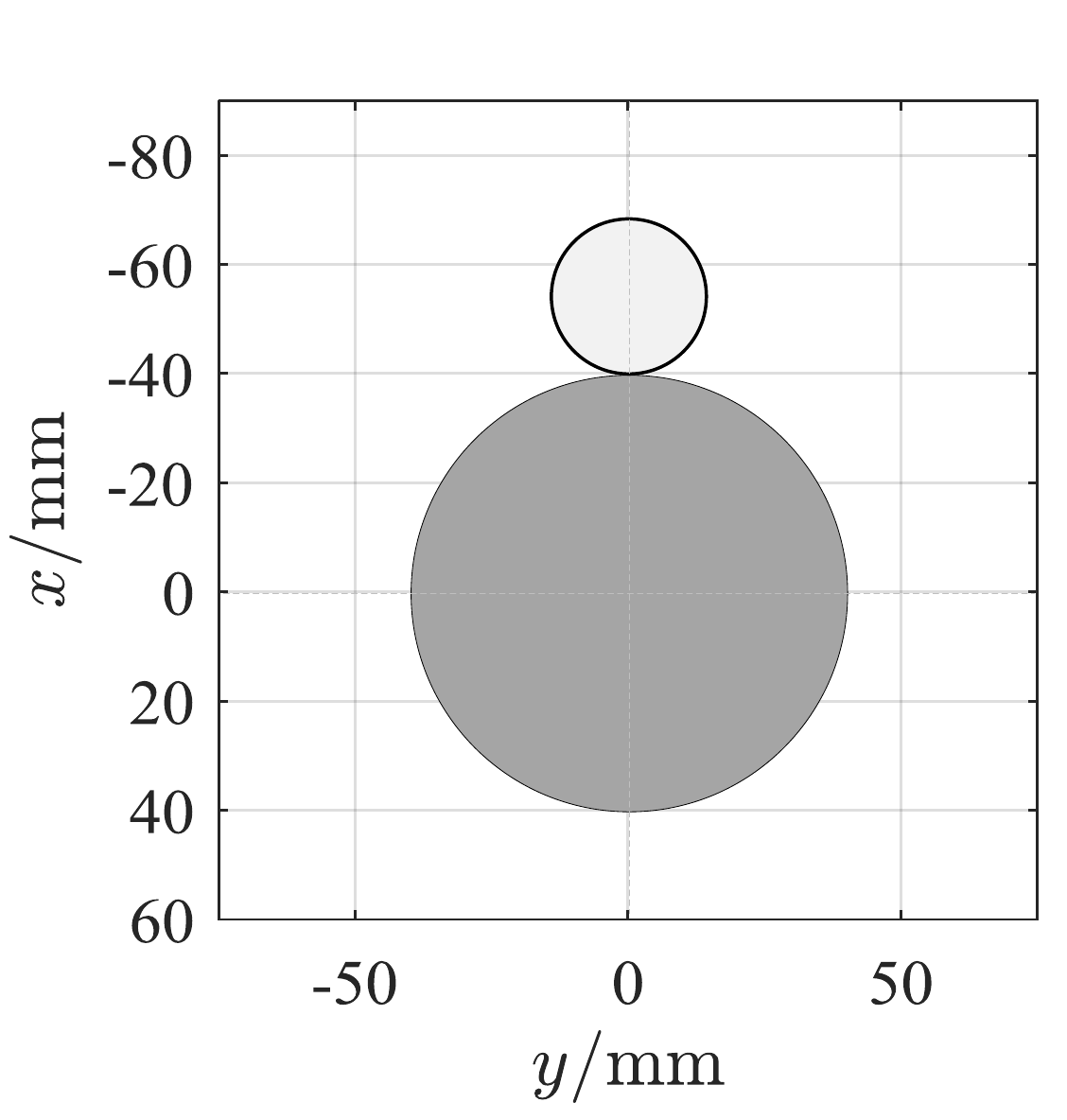}}\\
        \makebox[\columnwidth]{
        \subfloat[]{\includegraphics[height=0.35\linewidth] {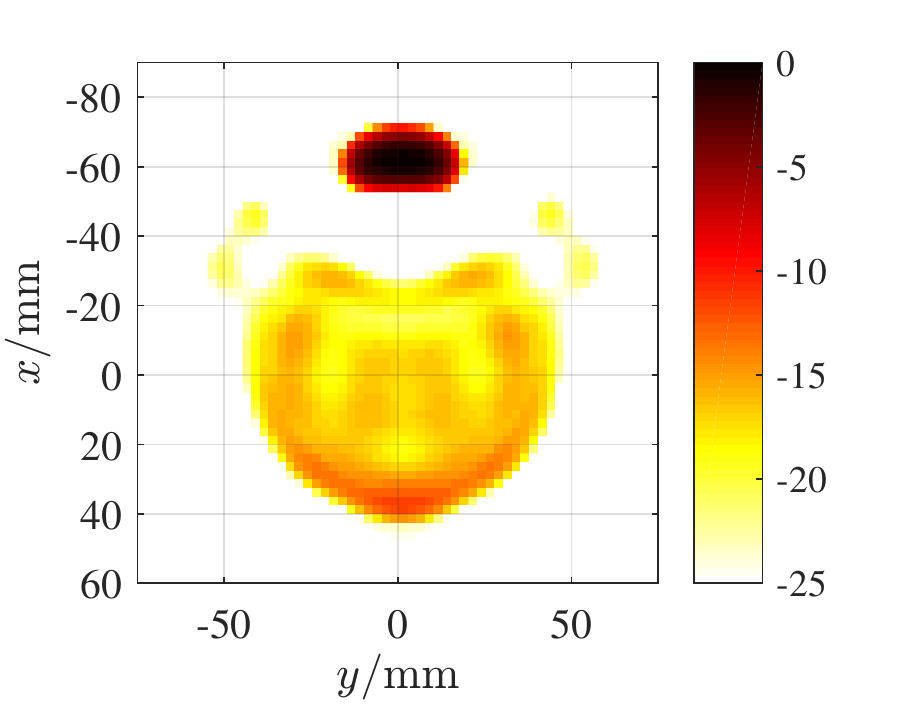}}%
        \subfloat[]{\includegraphics[height=0.35\linewidth] {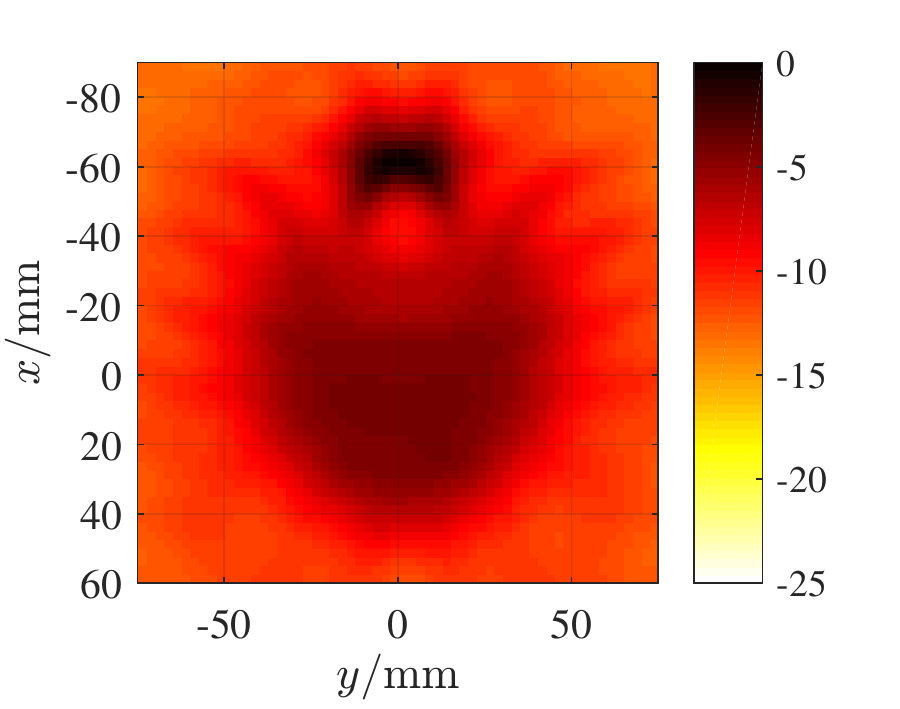}}}
        \caption{Scatterer geometry and its reconstructed shapes for the hybrid scatterers obtained by processing the multiple frequency data at 2 GHz, 3GHz, \dots, 8 GHz in Subsection~\ref{subsec.DieMet}: Scatterer geometry (a); The scatterer shape reconstructed by GMMV (b) and LSM (c).}
        \label{fig:DieMet}
    \end{figure}     

    In this subsection, we applied the proposed method to hybrid scatterers consisting of a foam circular cylinder (diameter = 80 mm, $\varepsilon_r = 1.45 \pm 0.15$) and a copper tube (diameter = 28.5 mm, thickness = 2 mm), which was considered in the \textit{FoamMetExtTM} data-set provided in the second opus of the Institut Fresnel’s database. We refer to \cite{geffrin2005free} for more description of the targets. The measurement configuration is the same with the one shown in Fig.~\ref{fig:DieMetConf}. In contrast to the \textit{FoamDieIntTM} data-set, this data-set is obtained using 18 transmitters, while other settings are kept the same. Specifically, the source antenna stays at the fixed location ($\theta = 0^\circ$), and the object is rotated to obtain different illumination incidences from 0$^\circ$ to 340$^\circ$ in steps of 20$^\circ$. 

    Let us first restrict the inversion domain to [$-90$, 60] $\times$ [$-75$, 75] mm$^2$ and discretize this domain with 2.5 $\times$ 2.5 mm$^2$ grids. The multi-frequency data at 7 frequencies, 2 GHz, 3 GHz, $\dots$, and 8 GHz, are jointly processed. The data matrix $\mF_i$ for LSM is a $360 \times 18$ matrix in which the data entries that are not available are replaced with zeros. Fig.~\ref{fig:DieMetCV} gives the normalized residual curves of the GMMV-based linear method, and the reconstructed images by both methods are shown in Fig.~\ref{fig:DieMet}. As we can see both the metallic cube and the circular foam cylinder can be well reconstructed by the GMMV-based linear method with high resolution, but for a slight part lost in between. In addition, one can also see from the GMMV image that the metallic cube obviously has larger intensity than the foam cylinder, showing a potential ability of distinguishing dielectric objects and metallic objects. In contrast, LSM shows a poor focusing ability in the hybrid scatterer case, indicating once again that the proposed GMMV-based linear method is able to achieve higher resolution image than LSM in this case.

\subsection{Computation Time}
    
    \begin{table}[!t]
        \renewcommand{\arraystretch}{1.3}
        \caption{Running times of the experimental examples}
        \label{tab.Num}
        \centering 
        \begin{tabular}{|l|>{\centering\arraybackslash}m{1.3cm}|>{\centering\arraybackslash}m{1.3cm}|>{\centering\arraybackslash}m{1.3cm}|}
        \hline
        Data-set    &   Frequency number    &   GMMV /s    &   LSM /s\\
        \hline
        \textit{twodielTM\_8f}      &  1   &   2.5      &   0.0145    \\ \hline 
        \textit{twodielTM\_8f}      &  4   &   12.7     &   0.0270    \\ \hline
        \textit{FoamDieIntTM}       &  5   &   3.8      &   0.0693    \\ \hline
        \textit{rectTM\_dece}       &  4   &   2.7      &   0.0225    \\ \hline
        \textit{uTM\_shaped}        &  4   &   41.0     &   0.0498    \\ \hline
        \textit{FoamMetExtTM}       &  7   &   15.6     &   0.0911    \\ \hline
        \end{tabular}
    \end{table}
    In this subsection, we discuss the computational complexity of the GMMV-based linear method. Since the sensing matrices can be computed (or analytically given for the experiments in homogeneous backgrounds) and stored beforehand, the GMMV-based linear method only involves a number of matrix-vector multiplications. The codes for reconstructing the contrast sources are written in MATLAB language. We ran the codes on a desktop with one Intel(R) Core(TM) i5-3470 CPU @ 3.20 GHz, and we did not use parallel computing. The running times of the GMMV-based linear method and LSM are listed in Table~\ref{tab.Num}, from which we see that, on one hand, all the reconstructions by the GMMV-based linear method require less than 1 minute (or even a couple of seconds for some examples); on the other hand, LSM shows overwhelmingly high efficiency in comparison to the GMMV-based linear method, because singular value decomposition (SVD) in LSM is done only once, then all of the indicator functions can be obtained simultaneously by several matrix-matrix multiplications. However, in view of the higher resolving ability of the proposed method, the extra computational cost is worth to pay. It is also worth mentioning that the proposed method is faster than the iterative shape reconstruction methods which solve the forward scattering problem in each iteration. In addition, parallel computing can be straightforwardly applied to the proposed method for acceleration.

\section{Conclusion}\label{sec.Conclusion}

    In this paper, a novel linear method for shape reconstruction based on the generalized multiple measurement vectors (GMMV) model is proposed. The sum-of-norm of the contrast sources at multiple frequencies was used for the first time as a regularization constraint in solving the electromagnetic inverse scattering problems. We applied this method to process 2-D transverse magnetic (TM) experimental data, and the results demonstrated that a regularized solution of the contrast sources by the sum-of-norm constraint is sufficient to recover the spatial profile of the non-sparse targets. Comparison results indicated that the GMMV-based linear method outperforms LSM in all the three cases of dielectric scatterers, convex and non-convex metallic scatterers, and hybrid scatterers in the shape reconstruction quality and the level of the sidelobes in the images. In view of the resolving ability and computational efficiency, the proposed method looks very promising in the application to three-dimensional imaging problems. Besides, the outcome of the GMMV-based linear method --- the contrast sources, can be directly used for quantitative imaging when the incident fields are known with a reasonable accuracy.    

\bibliographystyle{ieeetr}
\bibliography{mybib}

\end{document}